# A First Experiment at Interfacing Crystal Plasticity and Continuum Dislocation Dynamics

Khaled SharafEldin[a], Peng Lin[a,1], Vignesh Vivekanandan[a], Gustavo M. Castelluccio[b], Benjamin Anglin[c], Anter El-Azab[a]

[a] Purdue University, West Lafayette, IN 47907, USA

[b] Cranfield University, Bedfordshire MK43 0AL, UK

[c] Naval Nuclear Laboratory, West Mifflin, PA 15122, USA

**Abstract:**

A computational approach has been developed for the analysis of the properties of 3D dislocation substructures generated by the vector density continuum dislocation dynamics (CDD), within the framework of crystal plasticity. In the CDD framework, the dislocation density on the individual slip systems is represented by vector fields with a unique dislocation line direction at each point in space. The evolution of these density fields is governed by a set of transport-reaction equations coupled with crystal mechanics. This detailed picture of the dislocation system enables mesoscale plasticity simulations based on dislocation properties at the lattice level. In the current work, a computational approach based on streamline construction is proposed to obtain the statistical properties of the dislocation substructures generated by CDD. Streamlines are obtained by travelling along the tangent of the vector density and velocity fields of the dislocation system, which enables us to construct the dislocation lines and their paths in the deformed crystal in 3D. The streamlines are computed by numerical integration of a set of partial differential equations for the parameterized tangent fields. Here, we use this approach to extract microstructure parameters from the CDD simulations that are relevant to substructure-sensitive crystal plasticity models. These parameters include the mean free path and average mobile dislocation segment length, as well as the dislocation wall volume fraction, together with the corresponding distributions. The results show that, past a short initial rise during monotonic loading, both the mobile dislocation segment length and dislocation mean free path decrease with the applied strain, which is consistent with the models used in the crystal plasticity literature. At all strains, log-normal distributions were fit to the data for the dislocation free path and mobile segment length.

[1] Currently with Beihang University, Beijing, 100191, China.

## 1. Introduction

The mechanical response of the metals that deform by dislocation mechanisms is an emergent effect representing the complex and hierarchical response of dislocations at both the lattice and mesoscopic scales. At the atomic scale, the stress required to move dislocations depends on the lattice nature and composition. At the mesoscale, the collective behavior of dislocations governs strain hardening, as long-range forces and reactions drive their organization into substructures that control the collective dynamics. The correlation between substructure formation and strain hardening was highlighted in numerous studies such as, Hansen and Huang (1998) and Hansen et al. (2006), which is attributed to the impediment of mobile dislocations via both short-range and long-range interactions. Mughrabi's seminal study (1983) elucidated the role of dislocation substructures in driving cyclic hardening and introduced the wall-channel composite model. In this model, relatively harder wall regions with high dislocation density predominantly experience elastic strain, while softer channel regions contribute most of the plastic strain through the unobstructed motion of mobile dislocations. The internal back-stresses generated as a result of the heterogeneous dislocation structure have been widely used to reproduce the Bauschinger effect.

Estrin et al. (1998) extended the composite model to describe the mechanical response under large monotonic strains by quantifying the volume fraction of sessile dislocation walls. The authors proposed that the wall fraction, $(f_w)$, decreases with increasing plastic shear strain, $(\gamma)$, according to,

$$f_w = f_\infty + (f_0 - f_\infty)\exp\left(\frac{-\gamma}{\bar{\gamma}}\right), \tag{1}$$

where $f_\infty$ denotes the wall volume fraction at sufficiently large strain, $f_0$ is the peak wall volume fraction, and $\bar{\gamma}$ is a material constant governing the decay rate of $f_w$. The experimentally observed reduction in wall fraction with plastic shear strain correlates with a corresponding decrease in the strain-hardening rate. Within the framework of crystal plasticity, Sauzay (2008) employed the composite model to derive an analytical back-stress formulation based on the Eshelby inclusion formalism under the assumptions of infinitesimal strain and linear elasticity. Building on this approach, Castelluccio and McDowell (2017) further proposed a mesoscale-sensitive crystal plasticity model in which dislocation substructures are explicitly parameterized to predict the mechanical response of face-centered cubic (FCC) metals under cyclic loading. In their formulation, the crystallographic back-stress rate, $(\dot{B}^{\alpha})$, is taken to be proportional to the crystallographic shear rate, $(\dot{\gamma}^{\alpha})$, according to,

$$\dot{B}^{\alpha} \propto \frac{f_w}{1 - f_w}\,\dot{\gamma}^{\alpha}, \tag{2}$$

in which the wall fraction decreases with increasing the plastic shear range $(\Delta\gamma^{max}/2)$, following,

$$f_w = f_\infty + (f_0 - f_\infty)\exp\left(\frac{-\Delta\gamma^{max}/2}{g_p}\right), \tag{3}$$

which is similar to the formulation proposed by Estrin et al. (1998). In Eq.(3), $g_p$ is a fitting parameter. In addition, Dindarlou and Castelluccio (2022) incorporated the similitude principle by Sauzay and Kubin (2011)

$$d_{struc} \propto \tau^{-1} \tag{4}$$

to relate the characteristic wall spacing, $(d_{struc})$, to the resolved shear stress, $(\tau)$. In this framework, $d_{struc}$ and the mobile dislocation segment length, $(l_{mobile})$, are considered to be equivalent $(d_{struc} \cong l_{mobile})$. $l_{mobile}$ is linked to the dislocation mean free path, $(l_{free})$, through a mesoscale parameter,

$$\eta = l_{free}/l_{mobile}, \tag{5}$$

The ratio $(\eta)$ between the dislocation mean free path and the mobile dislocation segment length presented in Eq. (5) which infers the dislocation substructure at the mesoscale crystal plasticity model. The plastic shear is then used to conditionally determine the corresponding $\eta$ that represents the dislocation structure (Castelluccio and McDowell, 2017). Using Eq. (4), the mean free path can be calculated by assuming that walls are impenetrable. Parameterization is required in order to predict the evolution of $f_w$, $l_{free}$, and $l_{mobile}$ for any crystallographic orientation and plastic strain, which was obtained from the analysis of extensive transmission electron microscopy (TEM) images.

The framework developed by Castelluccio and McDowell (2017) coupled with proper quantification of mesoscale parameters ($\mathrm{f_w}$, $\mathrm{l_{free}}$, and $\mathrm{l_{mobile}}$), has been recently extended to monotonic loading (Dindarlou and Castelluccio, 2022). Here, they further assumed the similitude principle in Eq. (4), and proposed that $\mathrm{f_w}$ assimilate to a phase transformation evolution and follows,

$$\mathrm{f_w} = \mathrm{f_\infty}\,\mathrm{erf}\left(\frac{\gamma^{eq}}{\gamma_0^{eq}}\right), \qquad (6)$$

in which $\gamma^{eq}$ which corresponds to the weighted summation of the accumulated shear on the most active slip system and its corresponding Hirth and Cottrell slip systems. Also, $\gamma_0^{eq} \sim 10$ is the inverse of the typical equivalent shear strain upon reaching Stage III hardening at which point $\mathrm{f_w}$. The morphology of substructures, which depends on the crystallographic orientation, is well suited to estimate the anisotropic mean free path of mobile dislocation as highlighted by Devincre et al. (2008). Additionally, the substructure-sensitive models have the ability to directly compare mesoscale simulations against transmission electron microscopy (TEM) observations. In contrast, alternative approaches, such as those based on dislocation junction formation (Grilli et al., 2018) or on geometrically necessary dislocations (GND) (Jiang et al., 2015), which both depend on dislocation density measurements that are often subject to high uncertainty, making local-scale validation more challenging. The well-known issue of non-unique crystal plasticity parameters (Ashraf and Castelluccio, 2021) can therefore be alleviated by developing models that undergo independent validation across multiple length scales. Substructure-sensitive model assessment relies not only on matching the homogenized stress-strain response, but also on comparing predicted and observed wall spacing and wall volume fraction. Despite its demonstrated effectiveness, the composite model requires determining the volume fraction occupied by regions of highly localized dislocation density, a quantity that varies with loading path and crystallographic orientation with respect to the loading condition. As a result, extending its use to general loading conditions has been limited by the lack of TEM data that are systematically collected to serve the goal of informing crystal plasticity models of the nature just discussed.

Physics-informed, hierarchal frameworks that enable the study of deformation across multiple length and time scales provide a powerful route for predicting material behavior from first principles. Such approaches can accelerate materials optimization for various applications, reducing the reliance on experimental datasets to only the initial exploration and final verification steps. This advantage has become increasingly important with the rising demand for high-fidelity, physics-rich datasets that can be utilized in developing machine learning models or in-depth microstructural analysis (Bertin et al., 2024; Bishara et al., 2022; Gu et al., 2024; SharafEldin et al., 2025). Among these frameworks, discrete dislocation dynamics (DDD) has been extensively applied to investigate cooperative dislocation behavior at nano- and micron-scales (Csikor et al., 2007; Cui et al., 2014; Groh and Zbib, 2009; Stricker et al., 2018; Stricker and Weygand, 2015). In the context of dislocation substructures, Lavenstein and El-Awady (2019) used DDD to examine the localization of edge dislocations within sessile walls, while Wu and Zaiser (2021) parameterized the evolution of wall spacing and wall thickness from DDD simulations, both quantities directly relevant to crystal plasticity formulations.

Despite its predictive capabilities, DDD is computationally intensive, which limits the spatial and temporal scales accessible to simulation and can influence the predictability of the resulting substructure patterns. As a result, effectively bridging DDD extracted microstructural features to engineering-scale crystal plasticity models remains a significant challenge. Studying the collective behavior of dislocations using continuum dislocation dynamics (CDD) provides an effective alternative to DDD which drastically changes the computational scaling dependance on the number of dislocation segments, to a fixed cost of the discretized simulation domain. Continuum representation of dislocations was established by considered by Nye (1953) and Kröner (1959) using a dislocation density tensor field $\boldsymbol{\alpha}$ whose evolution is governed by an equation of the form: $\dot{\boldsymbol{\alpha}} = \boldsymbol{\nabla} \times (\mathbf{v} \times \boldsymbol{\alpha})$ proposed by Mura (1963) and Kosevich (1965), with $\mathbf{v}$ being the velocity vector. Multitude of CDD frameworks have been developed, see Arsenlis et al. (2004), Reuber et al. (2014), Leung et al. (2015), Hochrainer (2015), and Monavari and Zaiser (2018), which are reviewed in El-Azab and Po (2018) and Anderson et al. (2022). Current CDD models that of relevance to making a connection with crystal plasticity is the vector-density based continuum dislocation dynamics (V-CDD) developed by Xia and El-Azab (2015), Xia et al. (2016), Lin and El-Azab (2020), Lin et al. (2021a), Lin et al. (2021b) and the more recent treatment for dislocation reactions Vivekanandan et al. (2021) and Vivekanandan et al. (2023).

In analyzing dislocation dynamics datasets to extract information relevant to crystal plasticity the complexity of the dislocation network becomes a challenge. As such, there is a growing demand for suitable algorithms of analysis of the dislocation networks. A data-mining approach has been proposed by Song et al. (2021) to express the strain energy density of a dislocation system as a function of dislocation density field variables and as a function of the coarse graining voxel size. A systematic coarse-graining analysis of DDD simulations has been presented recently (Akhondzadeh et al., 2020, 2021) to formulate a generalized Taylor relation and a generalized Kocks-Mecking model. Machine learning models also have been used to deduce the dislocation characteristics from large sets of dislocation dynamics simulations (Rafiei et al., 2020; Steinberger et al., 2019; Yang et al., 2020).

The current work considers the CDD formulation by Vivekanandan et al. (2021) to establish a data-mining approach to characterize the mesoscale quantities ($f_w$, $l_{free}$, and $l_{mobile}$) relevant to crystal plasticity using CDD simulations. The morphology of the dislocation substructures generated by CDD is then analyzed using streamlines along the dislocation density and velocity fields. Together with the magnitude of the dislocation velocity, these streamlines enable us to extract the quantities $f_w$, $l_{free}$, and $l_{mobile}$ that can be compared with experiments. The results demonstrate that CDD is capable of informing higher level crystal plasticity models and bridge effectively with continuum models.

## 2. Methodology

### 2.1. The vector-density based continuum dislocation dynamics

In CDD, it is assumed that dislocations locally have a unique line direction at a given point in space, so the dislocation density vector can be viewed as a dislocation bundle. The assumption is satisfied by choosing a relatively small mesh. Following Nye (1953) and Kröner (1959), we express dislocation density tensor, $\boldsymbol{\alpha}$, as

$$\boldsymbol{\alpha} = -\nabla \times \boldsymbol{\beta}^{\mathrm{p}}, \tag{7}$$

with $\boldsymbol{\beta}^{\mathrm{p}}$ being the plastic distortion tensor. Both tensors can be decomposed into slip system contributions,

$$\boldsymbol{\alpha} = \sum_{\mathrm{k}} \boldsymbol{\alpha}^{(\mathrm{k})}, \tag{8}$$

$$\boldsymbol{\beta}^{\mathrm{p}} = \sum_{\mathrm{k}} \boldsymbol{\beta}^{(\mathrm{k})}, \tag{9}$$

where $\mathrm{k} \in \mathcal{K}$ is a slip system index and $\mathcal{K}$ the set of slip systems. As dislocations move, the plastic distortion evolves and its time rate of change follows Orowan's law,

$$\dot{\boldsymbol{\beta}}^{\mathrm{p}^{(\mathrm{k})}} = -\mathbf{v}^{(\mathrm{k})} \times \boldsymbol{\alpha}^{(\mathrm{k})}, \tag{10}$$

where $\mathbf{v}^{(\mathrm{k})}$ is the dislocation velocity on slip system k. Here, we assume the resolution is sufficiently fine such that the line direction of a dislocation segment is unique at each point in the crystal. As such, the direction of the dislocation velocity $\mathbf{v}^{(\mathrm{k})}$, which is perpendicular to the dislocation line is uniquely defined. The dislocation density vector $\boldsymbol{\rho}^{(\mathrm{k})}$ is used to represent the dislocation density and line direction at all points. The relation between the dislocation density vector $\boldsymbol{\rho}^{(\mathrm{k})}$ and the dislocation density tensor $\boldsymbol{\alpha}^{(\mathrm{k})}$ is stated as

$$\boldsymbol{\alpha}^{(\mathrm{k})} = \boldsymbol{\rho}^{(\mathrm{k})} \otimes \mathbf{b}^{(\mathrm{k})}, \tag{11}$$

where $\mathbf{b}^{(\mathrm{k})}$ is the Burgers vector of dislocations on slip system k.

Combining Eqs. (7) through (11), the evolution equation for the dislocation density vector $\boldsymbol{\rho}^{(\mathrm{k})}$ in the absence of reactions and cross slip can be formulated as (Xia and El-Azab, 2015),

$$\dot{\boldsymbol{\rho}}^{(\mathrm{k})} = \nabla \times \left(\mathbf{v}^{(\mathrm{k})} \times \boldsymbol{\rho}^{(\mathrm{k})}\right), \tag{12}$$

where the Burgers vector was dropped. In addition to the evolution of the dislocation density vector via dislocation glide described by Eq. (12), dislocation reactions among different slip systems also contribute to its evolution. Therefore, additional terms must be added to Eq. (12) to account for cross slip, using the method described by (Vivekanandan et al., 2023, 2022) replacing the time-series analysis by Xia et al. (2016), collinear annihilation and junction reactions (Lin and El-Azab, 2020; Vivekanandan et al., 2021). Coupling these dislocation reactions with dislocation transport, the final form controlling the evolution of dislocations in CDD is

$$\dot{\boldsymbol{\rho}}^{(\mathrm{k})} = \nabla \times \left(\mathbf{v}^{(\mathrm{k})} \times \boldsymbol{\rho}^{(\mathrm{k})}\right) - \dot{\boldsymbol{\rho}}_{\mathrm{cs}}^{(\mathrm{k,l})} + \dot{\boldsymbol{\rho}}_{\mathrm{cs}}^{(\mathrm{l,k})} - \dot{\boldsymbol{\rho}}_{\mathrm{col}}^{(\mathrm{k,l})} - \dot{\boldsymbol{\rho}}_{\mathrm{g}}^{(\mathrm{kl,m})} + \dot{\boldsymbol{\rho}}_{\mathrm{g}}^{(\mathrm{lm,k})} - \dot{\boldsymbol{\rho}}_{\mathrm{L}}^{(\mathrm{kl,m})} - \dot{\boldsymbol{\rho}}_{\mathrm{H}}^{(\mathrm{kl,m})} \tag{13}$$

where $\dot{\boldsymbol{\rho}}_{\mathrm{cs}}^{(\mathrm{k,l})}$ and $\dot{\boldsymbol{\rho}}_{\mathrm{cs}}^{(\mathrm{l,k})}$ are rate of change of dislocation density from and to slip system k due to cross slip, $\dot{\boldsymbol{\rho}}_{\mathrm{col}}^{(\mathrm{k,l})}$ is the rate of change of dislocation density due to collinear annihilation, $\dot{\boldsymbol{\rho}}_{\mathrm{g}}^{(\mathrm{kl,m})}$

accounts for the rate of change of dislocation density due to all glissile junctions with $\boldsymbol{\rho}^{(k)}$ as reactant, and $\dot{\boldsymbol{\rho}}_{g}^{(lm,k)}$ accounts for the rate of change of dislocation density due to all glissile junctions with $\boldsymbol{\rho}^{(k)}$ as product. The last two terms in Eq. (13) account for Lomer-Cottrell junction and Hirth junction formation. In this work, Eq. (13) only considers cross slip, collinear annihilation and glissile junctions to capture the dislocation density evolution. The effect of Lomer-Cottrell junction and Hirth junction to dislocation motion is considered via a Taylor law as described later (Eq. (21)).

The velocity field is required to solve Eq. (13). In the current model, the dislocation velocity is estimated by evaluating the internal long-range stress field from which the Peach-Koehler force for each slip system is evaluated then used to calculate the corresponding velocity via a dislocation mobility law. The long-range stress of the dislocations is calculated by solving the eigenstrain boundary value problem:

$$\begin{cases} \nabla \cdot \boldsymbol{\sigma} = 0 & \text{in V} \\ \boldsymbol{\sigma} = \mathbb{C} : (\nabla \mathbf{u} - \boldsymbol{\beta}^{p}) & \text{in V} \\ \mathbf{u} = \bar{\mathbf{u}} & \text{on } \partial V_{u}' \\ \mathbf{n} \cdot \boldsymbol{\sigma} = \bar{\mathbf{t}} & \text{on } \partial V_{\sigma} \end{cases} \tag{14}$$

where $\boldsymbol{\sigma}$ is the Cauchy stress, $\mathbb{C}$ is the symmetric, fourth rank elastic tensor, $\mathbf{u}$ is the displacement field, $\boldsymbol{\beta}^{p}$ is the plastic distortion, $\mathbf{n}$ is the unit normal to the traction boundary, $\partial V_{\sigma}$ and $\partial V_{u}$, respectively, are the parts of the boundary over which the traction $\bar{\mathbf{t}}$ and the displacement $\bar{\mathbf{u}}$ boundary conditions are prescribed. The plastic distortion $\boldsymbol{\beta}^{p}$ is updated by the decomposition $\boldsymbol{\beta}^{p} = \nabla \mathbf{z} - \boldsymbol{\chi}$, with $\nabla \mathbf{z}$ and $\boldsymbol{\chi}$ being the compatible and incompatible parts of $\boldsymbol{\beta}^{p}$, respectively (Lin et al., 2021b). It has been shown that updating $\boldsymbol{\beta}^{p}$ by field dislocation mechanics is more accurate than directly integrating Orowan's equation (Lin et al., 2021b). These two components of the plastic distortion are governed by the following boundary value problems:

$$\begin{cases} \nabla \times \boldsymbol{\chi} = \sum_{k} \boldsymbol{\rho}^{(k)} \otimes \mathbf{b}^{(k)} & \text{in V} \\ \nabla \cdot \boldsymbol{\chi} = 0 & \text{in V} \\ \mathbf{n} \cdot \boldsymbol{\chi} = 0 & \text{on } \partial V \end{cases} \tag{15}$$

and,

$$
a\begin{cases}\nabla\cdot\nabla\dot{\mathbf{z}}=\nabla\cdot\sum_{\mathrm{k}}(-\mathbf{v}^{(\mathrm{k})}\times\boldsymbol{\rho}^{(\mathrm{k})}\otimes\mathbf{b}^{(\mathrm{k})}) & \text{in V}\\ \mathbf{n}\cdot\nabla\dot{\mathbf{z}}=\mathbf{n}\cdot\sum_{\mathrm{k}}(-\mathbf{v}^{(\mathrm{k})}\times\boldsymbol{\rho}^{(\mathrm{k})}\otimes\mathbf{b}^{(\mathrm{k})}) & \text{on }\partial\mathrm{V}\\ \dot{\mathbf{z}}=\dot{\mathbf{z}}_{\mathrm{o}}(\text{arbitrary value}) & \text{at one point in V}\end{cases}. \quad (16)
$$

Here, V is the simulation domain with boundary $\partial\mathrm{V}$. The decomposition above is similar in spirit to the field dislocation mechanics approach by (Acharya and Roy, 2006; Roy and Acharya, 2006) who rather apply the decomposition to the elastic distortion to fix the elastic solution.

The dislocation velocity vector in Eq. (13) is expressed in the form

$$
\mathbf{v}^{(\mathrm{k})}=\mathrm{v}^{(\mathrm{k})}\boldsymbol{\eta}^{(\mathrm{k})} \quad (17)
$$

with $\mathrm{v}^{(\mathrm{k})}$ being the scalar velocity and $\boldsymbol{\eta}^{(\mathrm{k})}$ a unit vector in the direction of dislocation motion determined by the slip plane normal $\mathbf{m}^{(\mathrm{k})}$ and the dislocation line direction $\boldsymbol{\xi}^{(\mathrm{k})}=\boldsymbol{\rho}^{(\mathrm{k})}/\rho^{(\mathrm{k})}$ (Lin et al., 2021b),

$$
\boldsymbol{\eta}^{(\mathrm{k})}=\mathbf{m}^{(\mathrm{k})}\times\boldsymbol{\xi}^{(\mathrm{k})}. \quad (18)
$$

The scalar velocity $\mathrm{v}^{(\mathrm{k})}$ is assumed to depend linearly on the resolved shear $\tau^{(\mathrm{k})}$ via the mobility law,

$$
\mathrm{v}^{(\mathrm{k})}=\mathrm{sgn}(\tau^{(\mathrm{k})})\frac{\mathrm{b}}{\mathrm{B}}[|\tau^{(\mathrm{k})}|-(\tau_0^{(\mathrm{k})}+\tau_{\mathrm{p}}^{(\mathrm{k})})] \quad (19)
$$

where $\mathrm{sgn}(\cdot)$ returns the sign of its argument, $\tau^{(\mathrm{k})}$ is the resolved shear stress, b is the magnitude of the Burgers vector, B is the drag coefficient, and $\tau_0^{(\mathrm{k})}$ and $\tau_{\mathrm{p}}^{(\mathrm{k})}$, respectively, are friction stress contributions due to lattice resistance and short-range interactions. In the above expression, $\mathrm{b}\tau^{(\mathrm{k})}$ corresponds to the magnitude of the Peach-Koehler force (Peach and Koehler, 1950), and the resolved shear stress itself is given by

$$
\tau^{(\mathrm{k})}=\mathbf{s}^{(\mathrm{k})}\cdot\boldsymbol{\sigma}\cdot\mathbf{m}^{(\mathrm{k})} \quad (20)
$$

where $\mathbf{s}^{(\mathrm{k})}=\mathbf{b}^{(\mathrm{k})}/\mathrm{b}$ is the unit slip direction. The resolved shear stress accounts for the long-range interactions between dislocations, dislocation-defect interactions, and boundary effects. The friction stress $\tau_0^{(\mathrm{k})}$ is the lattice threshold stress for dislocation motion (Hirth and Lothe, 1982; Hull and Bacon, 2011), while $\tau_p^{(k)}$ is the resistance due to short-range interactions (Deng and El-Azab, 2010; El-Azab, 2000; Hochrainer, 2016; Sandfeld and Zaiser, 2015). In this work, $\tau_p^{(k)}$ is modelled using the Taylor hardening law proposed by (Devincre et al., 2006; Franciosi et al., 1980; Kubin et al., 2008) as

$$\tau_p^{(k)} = \mu b \sqrt{\sum_{l}^{N_{ss}} a^{kl} \rho^{(l)}} \tag{21}$$

with $\mu$ being the shear modulus and $a^{kl}$ an interaction coefficient representing the average strength of the mutual interactions between slip systems k and l, between all slip systems ($N_{ss}$). As the collinear annihilation and glissile junction has been considered in Eq. (13), only sessile Lomer lock $(a^{kl} = 0.42)$ and Hirth junctions $(a^{kl} = 0.23)$ are considered in Eq. (21).

### 2.2. Mathematical description of streamlines in continuum dislocation dynamics

In 3D CDD models, dislocations are curved lines and interact with one another to form complicated dislocation networks. Consequently, both the dislocation density and dislocation line direction are available for analysis and visualization of the dislocation substructure. Unlike the case of DDD where dislocations are represented by discrete line segments, continuous dislocation density fields are used in CDD. In such a representation, the dislocation lines are smeared, and a technique is required to reconstruct dislocation lines from these field variables. Such a task can make use of the idea of streamlines, which are used in computational fluid dynamics to visualize fluid flow. Streamlines are families of curves that are locally tangent to the velocity vector in flowing fluids. In CDD, the dislocation density vector can thus be analyzed using streamlines.

Let $P(s) = \big(x(s), y(s), z(s)\big)$ represents the points on the streamlines of dislocations with coordinates $(x, y, z)$ parameterized with the scalar distance s along the line. This parameterization satisfies

$$\frac{dx}{ds}=\frac{\rho_x(x,y,z)}{\|\boldsymbol{\rho}\|},\quad \frac{dy}{ds}=\frac{\rho_y(x,y,z)}{\|\boldsymbol{\rho}\|},\quad \frac{dz}{ds}=\frac{\rho_z(x,y,z)}{\|\boldsymbol{\rho}\|}. \tag{22}$$

where $\boldsymbol{\rho}$ is the dislocation density vector and $\rho_x, \rho_y, \rho_z$ are its components. Given an initial point $P(0) = (x_0, y_0, z_0)$, it is possible to construct a series of points lying on the streamline passing through this initial point by solving the above set of differential equations. That is, we can reconstruct a series of points $P(s_1), P(s_2), P(s_3), \cdots, P(s_n)$ along the streamline. Doing so for numerous suitably chosen initial points for the density of a given slip system makes it possible to visualize the entire dislocation population on that slip system. Coloring the streamline based on the local density magnitude also enables both the dislocation lines and their local densities throughout the crystal one slip system at a time.

A typical dislocation density substructure contains regions of low density (channels) and regions of high density (walls). The structure length scale along the glide direction of dislocations can thus be used to characterize the dislocation mean free path while the distance along the streamline of the dislocation between pinning points can be associated with the length of the mobile segments. These length scales can directly inform crystal plasticity models (Castelluccio et al., 2018; Castelluccio and McDowell, 2017); see also (Devincre et al., 2008).

The length of a mobile dislocation segment is defined as the distance between two pinning points along a dislocation line. Pinning points are defined here to be the points at which dislocation velocity is smaller than a conveniently chosen small cutoff (tolerance), $v_{tol}$. Here it is taken to be $10^{-5}\mu m\ ns^{-1}$. The mobile dislocation segment length, $l_{mobile}$, is calculated using

$$\begin{gathered} l_{mobile}=\sum_{k=i}^{k=j-1}|s_{k+1}-s_k|\,,\quad \text{where} \\ \|\mathbf{v}(P(s_i))\|\le v_{tol},\quad \left\|\mathbf{v}\left(P(s_j)\right)\right\|\le v_{tol}, \text{and} \\ \|\mathbf{v}(P(s_k))\|>v_{tol},\quad i<k<j, \end{gathered} \tag{23}$$

where $s_k$ is the scalar distance along the density streamline at point k.

In order to fix the mean free path of dislocations, it will be required to analyze the velocity of the dislocations in the direction of their glide. For this purpose, the dislocation trajectories (path lines) can be determined by the streamlines of the dislocation velocity field over short periods of time centered about the time of interest. As discussed in Section 4.2.2, the dislocation velocity field does not significantly vary over the time that dislocations travel along their mean free path, which support the use of the dislocation movement path instead. If $Q(s) = (x(s), y(s), z(s))$ represents the points on the streamlines of dislocation velocity fields, these streamlines satisfy

$$\frac{dx}{ds} = \frac{v_x(x,y,z)}{\|\mathbf{v}\|}, \quad \frac{dy}{ds} = \frac{v_y(x,y,z)}{\|\mathbf{v}\|}, \quad \frac{dz}{ds} = \frac{v_z(x,y,z)}{\|\mathbf{v}\|}, \tag{24}$$

where $\mathbf{v}$ is the dislocation velocity vector and $v_x, v_y, v_z$ are its Cartesian components. So, the length of the dislocation free path, $l_{free}$, is calculated from

$$\begin{gathered} l_{free} = \sum_{k=i}^{k=j-1} |s_{k+1} - s_k|, \quad \text{where} \\ \|\mathbf{v}(Q(s_i))\| \leq v_{tol}, \quad \left\|\mathbf{v}\left(Q(s_j)\right)\right\| \leq v_{tol}, \text{and} \\ \|\mathbf{v}(Q(s_k))\| > v_{tol}, \quad i < k < j. \end{gathered} \tag{25}$$

while Eqs. (23) and (25) give the structural quantities of interest, their analyses require density-weighting in order to properly average them.

## 3. Numerical implementation

The numerical implementation of CDD consists of solving the stress equilibrium problem for the displacement field $\mathbf{u}$ given the plastic distortion field and solving the dislocation kinetics equations for all $\boldsymbol{\rho}^{(k)}$. The plastic distortion is updated using the dislocation information by updating its compatible ($\mathbf{z}$) and incompatible ($\boldsymbol{\chi}$) parts and solving two subsidiary boundary value problems. A staggered scheme (Lin et al., 2021b; Xia and El-Azab, 2015) is used to decouple the mechanics and dislocation kinetic problems, which amounts to solving the mechanical equilibrium given the plastic distortion, followed by an update for the dislocation evolution given the stress field. The solution scheme begins with the initialization of all slip systems with an initial dislocation density, which is then used to solve the field dislocation mechanics equations to obtain the corresponding plastic distortion $\boldsymbol{\beta}^p$. Following this step, the stress field is obtained by solving the stress equilibrium equation, which is then used to obtain the dislocation velocity based on the mobility law. The velocity obtained is then used to solve the kinetic equation to update the dislocation density evolution. The new dislocation configuration is used for the next timestep and the process is repeated until the average strain reaches the desired value. Two finite element methods are implemented to solve the overall problem. The mechanical equilibrium equations are solved using the Garlerkin method (Belytschko et al., 2014) and the dislocation transport equations are solved using the first order systems least squares (FOSLS) method (Jiang, 1998), which enables the incorporation of the divergence-free constraints of the dislocation densities. An implicit Euler method (Press et al., 2007; Xia and El-Azab, 2015) is used for time discretization. The time step is controlled by a Courant-Friedrichs-Lewy condition (Lax, 1973), which ensures that the fastest dislocation line bundle cannot move through an entire element in one timestep. The streamline calculation is done by an implicit Runge-Kutta method (Press et al., 2007).

## 4. Results and discussion

### 4.1. The evolution of dislocation loops represented by streamlines

In this section, a simple example is set up to illustrate the difference between two representations of the dislocation substructures obtained from the dislocation density fields. One is the commonly used scalar dislocation density, and the other is the streamline of the vector dislocation densities as proposed in this paper. The example shows the annihilation of two dislocation loops, which are initially located on two collinear slip systems of an FCC crystal, $(111)[0\bar{1}1]$ and $(\bar{1}11)[0\bar{1}1]$, and expands with a prescribed constant dislocation velocity, as shown in Figure 1. Since the dislocation lines have the same Burgers vector but opposite line directions at the intersection part, they annihilate with each other and form a larger dislocation loop across two slip planes. The evolution of dislocations on the two slip systems is tracked by solving the transport equations for the two dislocation density fields $\boldsymbol{\rho}^{(1)}$ and $\boldsymbol{\rho}^{(2)}$. To show the results in 3D, one commonly used method is to calculate the scalar dislocation density $\rho$, which is defined as the norm of the vector dislocation density $\|\boldsymbol{\rho}^{(1)}\| + \|\boldsymbol{\rho}^{(2)}\|$. The scalar dislocation density is shown in Figure 1(a) and Figure 1(c). Lower and higher dislocation regions are distinguished by the legend. Figure 1 demonstrates that the scalar dislocation density describes key features of the annihilation process but, by its nature, lacks the ability to show the directions of dislocation lines. To enhance our ability to perform analysis of the dislocation substructure, the streamlines of the dislocation density fields can be used as shown in Figure 1(b) and Figure 1(d). The arrows on the streamlines show the dislocation line directions as tangents of the streamlines. We remark that the streamlines are only used to show the directions of the curved dislocation lines, and the number of the streamlines should not be interpreted as the actual number of dislocation lines.

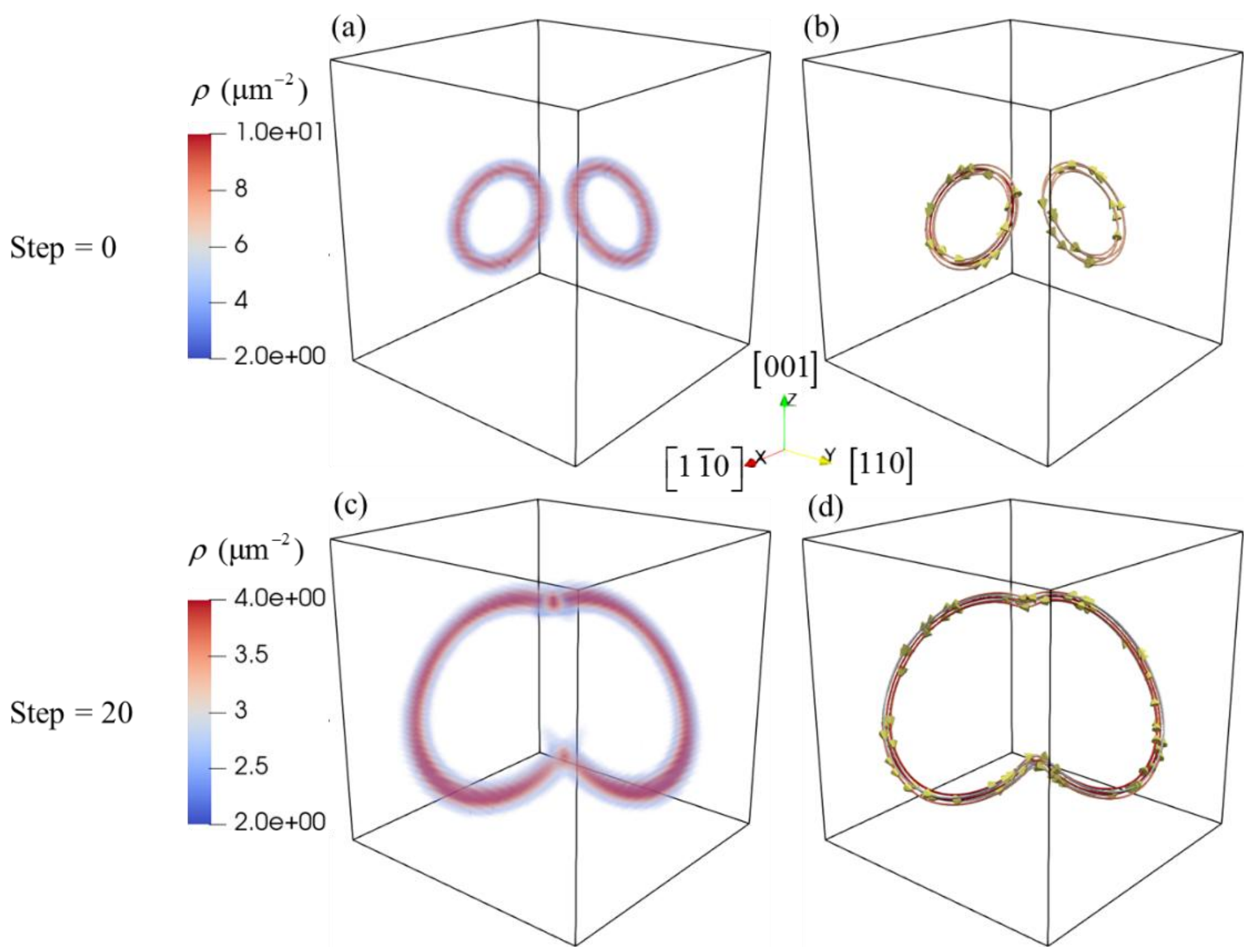

Figure 1. Collinear annihilation of two dislocation loops. (a), (c) Scalar density visualization. (b), (d) Streamline visualization.

### 4.2. The characteristics of dislocation substructures in a uniaxial loaded FCC crystal

In this section, a detailed analysis of the characteristics of dislocation substructures in a bulk simulation is performed using the streamlines of the dislocation density field. The simulation is performed in a 8 μm × 8 μm × 8.485 μm domain with all 12 slip systems of an FCC crystal considered. The edges of the simulation domain are along [110], $[\bar{1}10]$ and [001] crystallographic directions. The crystal is loaded along the [001] direction with a strain rate of 20 $s^{-1}$. Periodic boundary conditions are applied on all faces of the domain; the slip systems and their respective numbering are listed in Table 1.

Table 1: Slip systems of FCC crystals, with colormap. The color scheme below the table carries to the plots whenever slip system specific quantities are plotted.

| Slip system | 1 | 2 | 3 | 4 | 5 | 6 | 7 | 8 | 9 | 10 | 11 | 12 |
|---|---|---|---|---|---|---|---|---|---|---|---|---|
| Slip plane | $(111)$ | $(\bar{1}11)$ | $(\bar{1}11)$ | $(11\bar{1})$ | $(11\bar{1})$ | $(1\bar{1}1)$ | $(1\bar{1}1)$ | $(111)$ | $(111)$ | $(11\bar{1})$ | $(\bar{1}11)$ | $(1\bar{1}1)$ |
| Slip direction | $[0\bar{1}1]$ | $[0\bar{1}1]$ | $[101]$ | $[101]$ | $[011]$ | $[011]$ | $[\bar{1}01]$ | $[\bar{1}01]$ | $[\bar{1}10]$ | $[\bar{1}10]$ | $[110]$ | $[110]$ |

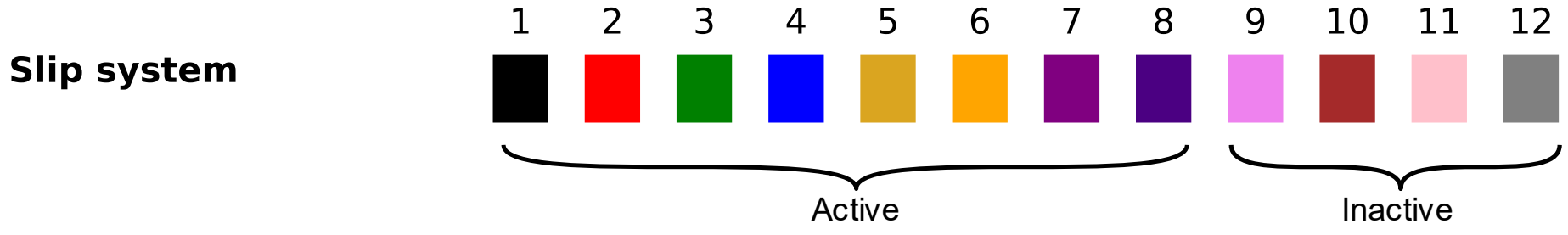

The material parameters for the simulations are: Young's modulus $\mathrm{E} = 207$ GPa, Poisson's ratio $\nu = 0.37$, Burgers vector $\mathrm{b} = 0.2489$ nm, drag coefficient $\mathrm{B} = 1.65 \times 10^{-5}\ \mathrm{Pa \cdot s}$. Initially, multiple dislocation loops are randomly placed in space, with periodic boundaries, on each slip system resulting in an initial dislocation density of $1.3\mu\mathrm{m}^{-2}$ distributed across all 12 slip systems by randomly placing 15 dislocation circular periodic bundles with initial random radius ranging from 5.5 μm to 9.5 μm. These parameters correspond to Ni (Olmsted et al., 2005).

The stress strain curve for this case is shown in Figure 2(a). A typical elastoplastic response is observed, with an initial elastic regime followed by strain hardening. The analysis of dislocation substructure in the following section is performed at different strain levels up to 2%. The evolution of total dislocation density is shown in Figure 2(b). The dislocation density increases nearly linearly from $1.3\mu\mathrm{m}^{-2}$ to $8.1\mu\mathrm{m}^{-2}$ up to a strain of 1.5%. Beyond this point, the total dislocation density exhibits a quasi-saturation behavior, suggesting a temporary stabilization of the microstructure. However, near 2% strain, the dislocation density appears to resume its linear growth rate. Figure 2(c) shows the plastic slip rate ($\dot{\gamma}$) carried out by each active slip system. As expected for uniaxial monotonic loading along $[001]$ direction, eight slip systems are activated, each having an identical Schmid factor. Ideally, this loading condition should yield the same slip response across all active systems. However, the results reveal a pronounced deviation from this ideal situation. Rather, we notice that the slip systems bifurcate into two distinct groups, one characterized by relatively higher plastic strain rates (slip system 1, 4, 5, and 8) and the other by lower rates (slip system 2, 3, 6, and 7). Despite this asymmetry, the average plastic shear rate across all slip systems remains nearly constant at approximately $6\mathrm{s}^{-1}$ under the present loading conditions. Figure 2(d) shows the slip system dislocation density evolution, which does not necessarily mirror the same slip system grouping previously observed in $\dot{\gamma}$. It is important to note that the dislocation density increase for inactive slip systems (slip system 9, 10, 11, and 12) is associated with dislocation storage through glissile junction mechanism, and not due to bow out or slip of dislocations, which is shown in Figure 2(c) where there is no inactive slip system contribution.

Experience with CDD simulations of face-centered cubic (FCC) materials show that different initial conditions cause variability in slip system activity with slip rates above and below the expected values based on Schmid factors alone. This behavior can be rationalized by the microstructural fluctuations and dislocation interactions associated with latent hardening, which break the ideal symmetry dictated by crystallography found in classical crystal plasticity works (Asaro and Rice, 1977; Franciosi and Zaoui, 1982). Current implementation of CDD which explicitly accounts for cross slip and junction formation (Vivekanandan et al., 2023), takes into consideration the non-slip Escaig stress in addition to Schmid stress and allows for dislocation storage in inactive slip systems through glissile junction formation. This slip free dislocation evolution has also been analyzed by DDD (Akhondzadeh et al., 2021).

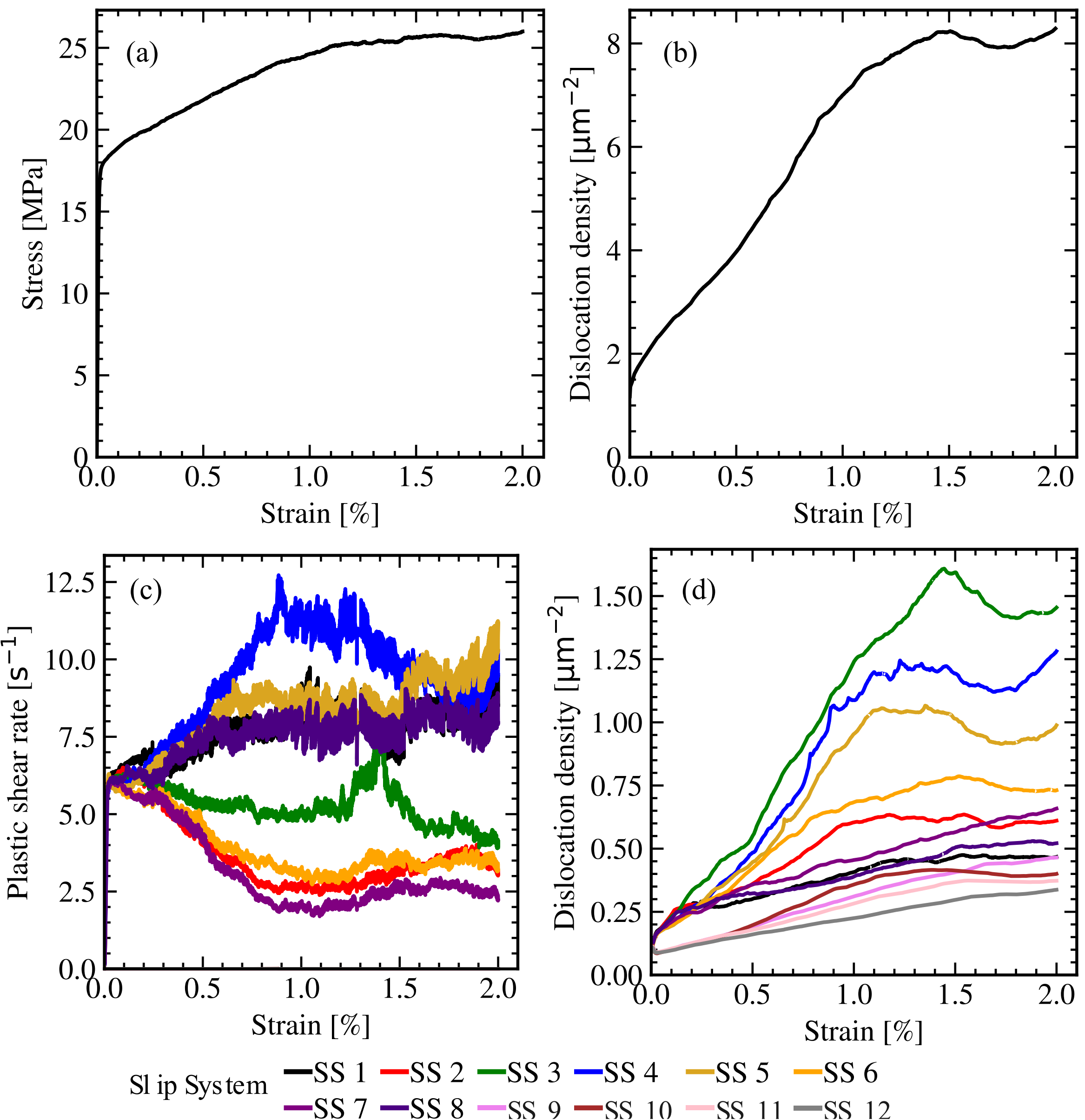

Figure 2. CDD prediction of (a) stress-strain curve, (b) evolution of total dislocation density, (c) plastic slip rate for each slip system, and (d) dislocation density evolution for each slip system.

The evolution of the overall dislocation density in the simulation domain is visualized at different strain levels (0.0%, 0.4%, 1.0%, and 1.6%) in Figure 3 showing the full domain (a-d) and a clipped view showing the $(1\bar{1}1)$ slip plane. The visualized range has an upper threshold equal to the $98^{th}$ percentile of the dislocation density distribution for improved visualization. Coarse dislocation-free equiaxed volumes are observed at 0.4% strain with approximately 2-4 μm diameter, this dislocation structure is separated by regions of high dislocation density. The regions of high dislocation density can be categorized into a very high dislocation density geometrically necessary boundaries (GNBs), and an intermediate density range that represents incidental dislocation boundaries (IDBs). At higher strain such as 1.0% and 1.6%, the famous cell structure develops within the spaces confined between the GNB structures. After 1%, the formed structures appear to be stable than the initial structure at lower strain. This observation was recently reported by Barkia et al. (2025) for 316L stainless steel where coarse structures initially formed, followed by cell block structures that appear to persist at larger strains in the uniform composition regions. Given the developed microstructure, it is expected to observe the streamline analysis to predominantly appear within the soft cell interior regions with low dislocation density, therefore, the mobile dislocation population and dislocation free paths are extracted and overlayed on top of the total dislocation density.

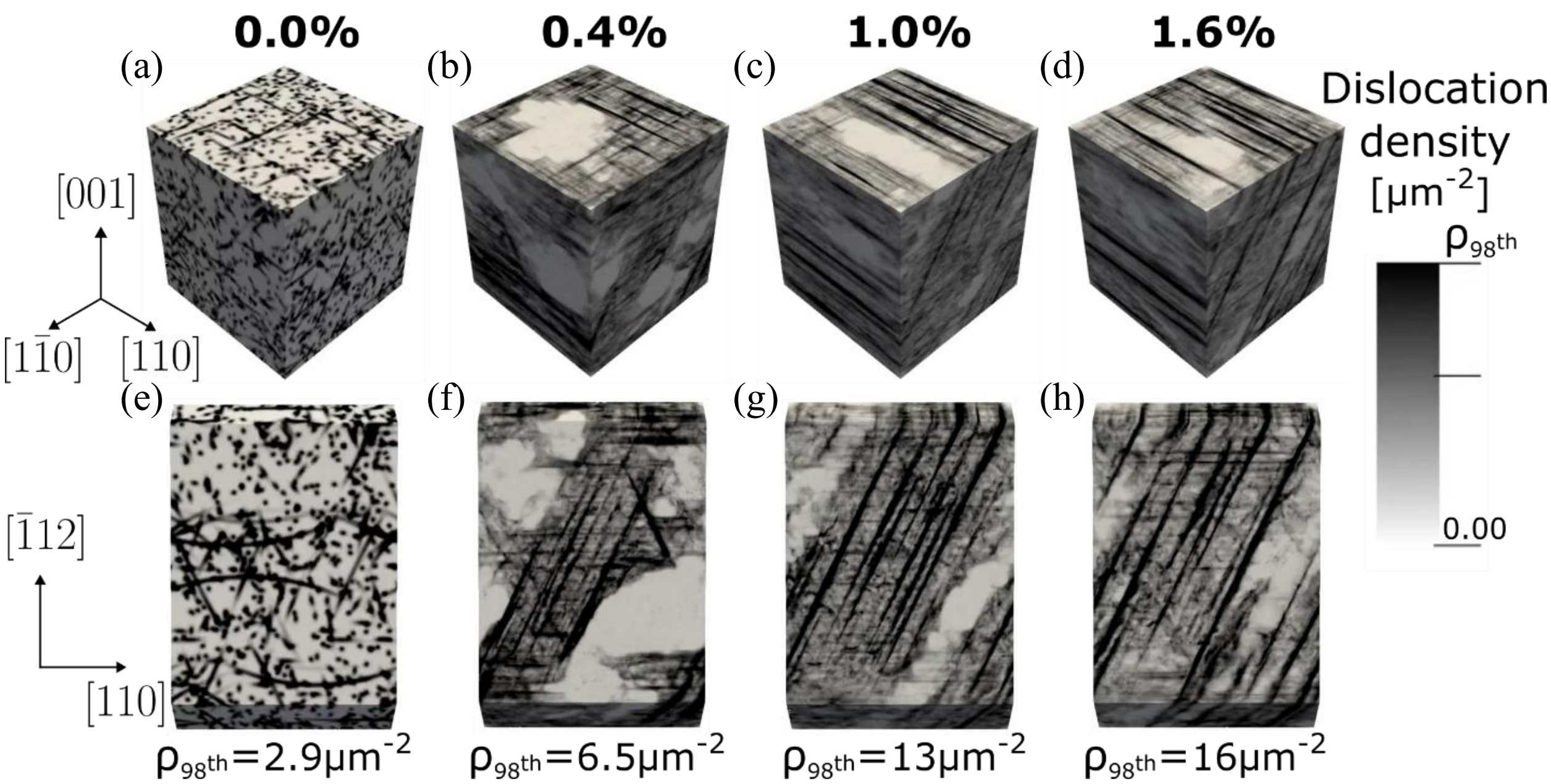

Figure 3. Distribution of overall dislocation density at different strain levels; (a,e) 0.0%, (b,f) 0.4%, (c,g) 1.0%, (d,h) 1.6%. An upper limit of 98$^{th}$ percentile of the dislocation density is imposed for a better visualization. (a-d) show the full simulation domain with dimensions 8μm×8μm×8.485μm along the $[1\bar{1}0]$, [110], and [001] directions, respectively, while (e-h) show the dislocation density on $(1\bar{1}1)$ section. The initial configuration is made of circular

loops, which evolves into initially isolated, large cells (0.4% images), before the mixture of GNBs and finer cells appear at larger strain values.

#### 4.2.1. Mobile dislocation segment length statistics

Accurate estimation of the mobile dislocation segment length is important in microstructure-based crystal plasticity models since it controls the bow out strength of dislocations pinned at cell walls (Castelluccio et al., 2018; Castelluccio and McDowell, 2017; Dindarlou and Castelluccio, 2022). In CDD models, the mobile segment length is computed by Eq. (22). The calculations are performed to identify the dislocation substructure at five different strains as shown in Figure 3. For each slip system, 1000 initial seed points are used to calculate the streamlines. The seed points are not chosen arbitrarily but are on the mesh points where dislocation velocities are sufficiently high to be considered as mobile. The streamline calculation is done in both directions from each seed point with 2000 points in each direction along the streamline.

The streamlines of dislocations on slip system $(111)[0\bar{1}1]$ at 0.4% and 1.6% strain are shown in Figure 4, in which the color of each streamline represents the local dislocation velocity. Parts (a,b) of the figure show the streamlines in the bulk, revealing line curvature. Dislocations with velocity above a velocity threshold of $\mathrm{v_{tol}} = 0.01\ \mathrm{m\ s^{-1}}$ are shown, which fall within the cell interior. This thresholding reveals clusters of streamlines within the cells with the peripheries of each cluster having lower velocities in comparison with cell interior. The shape of the mobile clusters at 0.4% strain is more equiaxed and bounded with relatively slow segments, influenced by the initial seed. At a higher strain of 1.6%, the cluster outline has evolved in an elongated shape as predicted by Dindarlou and Castelluccio (2022) for a similar crystal orientation. To further investigate the relationship between the dislocation substructure and the extracted mobile dislocation segments, Figure 4(c,d) overlay the total dislocation density over a $(111)$ section with the streamlines clipped slightly above that plane. The results shows that the majority of mobile segments form in the low dislocation density regions, which is typical of the soft cell interior. At 1.6% strain, the trailing arrays of dislocation segments appear to form between regions of high dislocation density, in a channel-like structure, forming narrow dislocation segment streamlines compared to the streamlines at 0.4% where the microstructure has not developed as much. Similar observations have been reported by Pham and Holdsworth (2014). This alignment in mobile dislocation segment distribution forming piled up groups can greatly influence the development of back stress within the dislocation structure.

The statistics of the mobile segment lengths can be analyzed by extracting the mobile dislocation segments from Figure 4 above the specified dislocation velocity threshold of $v_{tol} = 10^{-5}\ \mu m\ ns^{-1}$ ($0.01\ m\ s^{-1}$). A typical mobile dislocation segment configuration is illustrated in Figure 4(d), which corresponds to the same location as Figure 4(c). All immobile dislocation segments on the streamlines are removed.

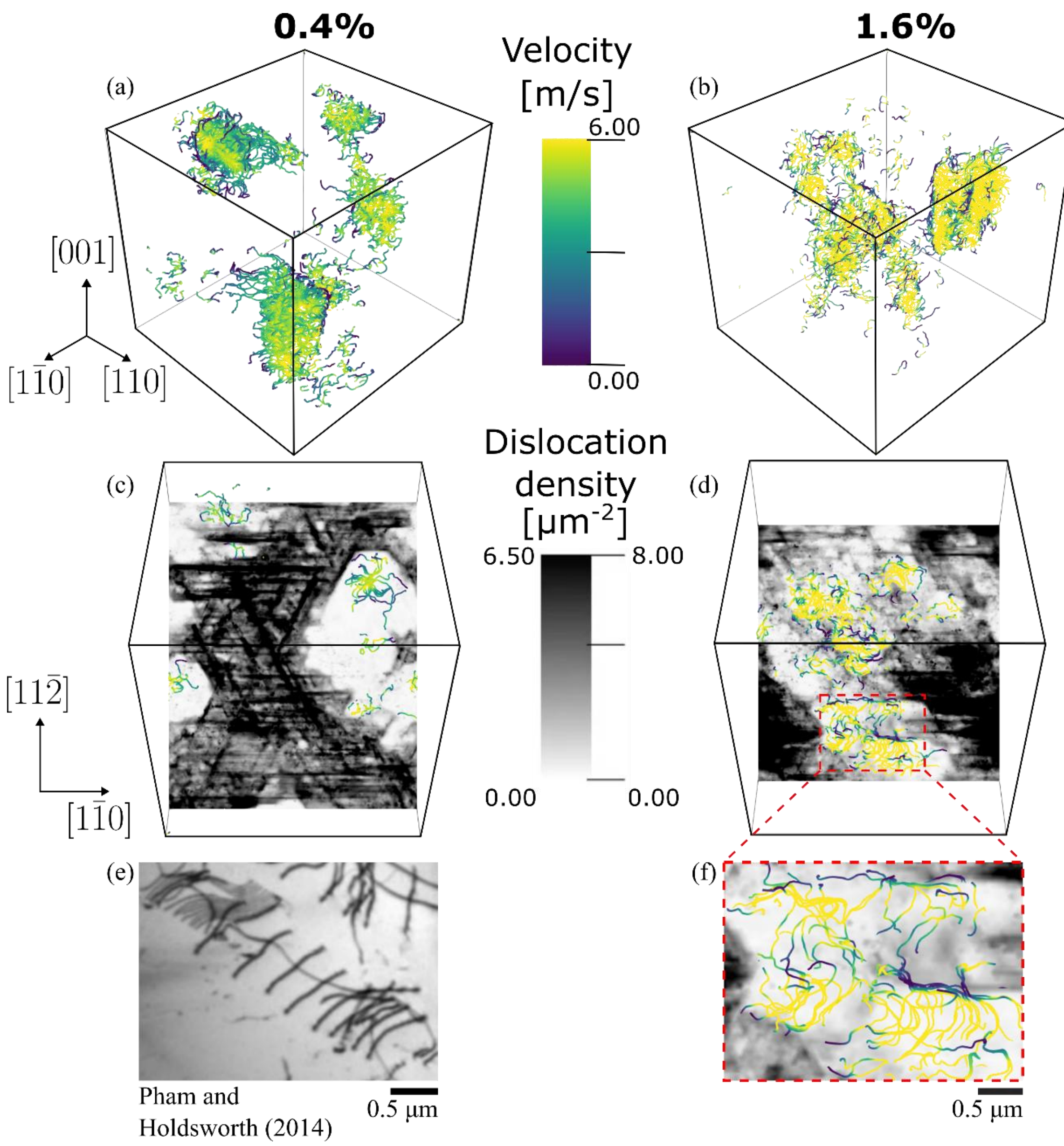

Figure 4. Streamlines of mobile dislocation segments for slip system $(111)[0\bar{1}1]$ at two strain levels: (a,c) 0.4%, and (b,d) 1.6%. Panels (a,b) show the full simulation domain with segments colored by velocity. Dislocations with velocities above the tolerence $v_{tol} = 10^{-5}$ $\mu m\ ns^{-1}$ are only shown, thus leaving our most of the sessile structures. Panels (c,d) overlay the total dislocation density on (111) plane with the mobile segments, which correlates the mobile streamline clusters with cell interiors. Segments with reduced mobility are pinned to the cell walls, whereas highly mobile segments are concentrated near the centers of dislocation cells. At 1.6% strain, arrays of trailing dislocations are highlighed in (f) consistent with experimental observations by Pham and Holdsworth (2014), adapted in panel (e).

The length of the mobile dislocation segment varies throughout the simulation domain. Hence, we first calculate the length of each mobile dislocation segment, $l^i_{mobile}$ and the dislocation density for that segment, $\rho^i$. A histogram is then constructed for the distribution of $l^i_{mobile}$ weighted by $\rho^i$. The reason for using $\rho^i$ as the weight is that the streamline only represents the line configuration but not the density. With an adequate number of streamlines, this counting method provides a histogram representative of the simulation domain volume. A higher local dislocation density at a given point indicates that there are more mobile dislocations associated with the visualized streamline at that point. The distributions of the mobile dislocation segment length for all eight active slip systems (for the given loading direction) at 1.0% strain are shown in Figure 5. The mobile dislocation segment length distribution follows a log-normal distribution,

$$P(l_{mobile}) = \frac{1}{l_{mobile}\sigma\sqrt{2\pi}} \exp\left(-\frac{(\ln(l_{mobile}) - \mu)^2}{2\sigma^2}\right) \quad (26)$$

as shown in Figure 5. The log-normal distribution indicates that $\ln(l_{mobile})$ has a normal distribution, with mean $\mu$ and standard deviation $\sigma$. This distribution is often useful to describe stochastic processes in material science (Gruber et al., 2008; Keller et al., 1999) and dislocation source lengths in DDD simulations (Shishvan and Van der Giessen, 2010); it has also been accounted for in crystal plasticity (Castelluccio and McDowell, 2017). Fitting the data to the log-normal distribution for each slip system, anisotropic behavior is observed across the slip systems by comparing fitting parameters.

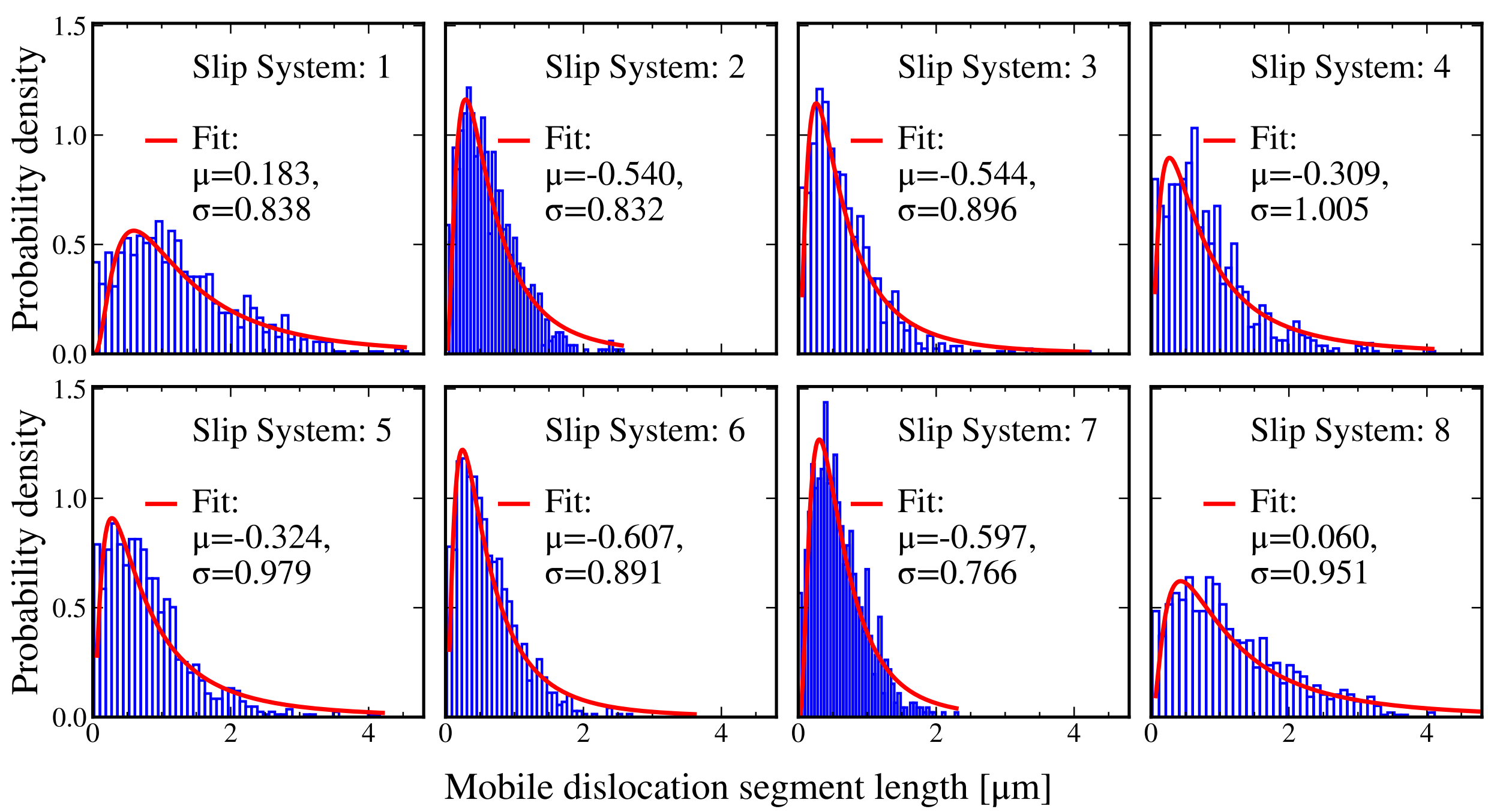

Figure 5. Distributions of mobile dislocation segment length for active slip systems at 1.0% strain. The data is plotted together with the log-normal fits, and the corresponding distribution parameters are listed on each panel.

The evolution of mobile dislocation segment length distribution as a function of strain for slip system $(111)[0\bar{1}1]$ is shown in Figure 6. The mobile dislocation segment length is observed to increase in length between 0.4% and 0.619%, afterwards, a clear decay of the length can be identified from 0.836% and beyond. By calculating the mean, $\mathrm{Mean}(l_{\mathrm{mobile}}) = \exp\left(\mu + \frac{\sigma^2}{2}\right)$, at 0.4%, the mean is 1.388 μm, which increases to 2.762 μm at 0.619% which ultimately decreases to 1.597 μm at 1.599% strain. This behavior can be attributed to the early development stage of microstructure evolution where dislocations initially glide with little to no obstructions, then the microstructure stabilize forming dislocation patterns similar to the evolution in Figure 3. The creation of stable structures hinders and subdivides the mobile segments, which would require additional stress to drive. The mutual pinning of dislocations at higher strains is also associated with the total dislocation density evolution in Figure 2. Forest hardening theories which predict the distance between the pinning points to vary as $1/\sqrt{\rho}$ where ρ is the total dislocation density (Baird and Gale, 1965; Kocks and Mecking, 2003; Schoeck and Frydman, 1972). Since the total dislocation density increases with strain, the distance between the pinning points will decrease.

Hence, one can expect the mobile dislocation segment length which is defined as the length of dislocation between two pinning points along the streamline to decrease with strain.

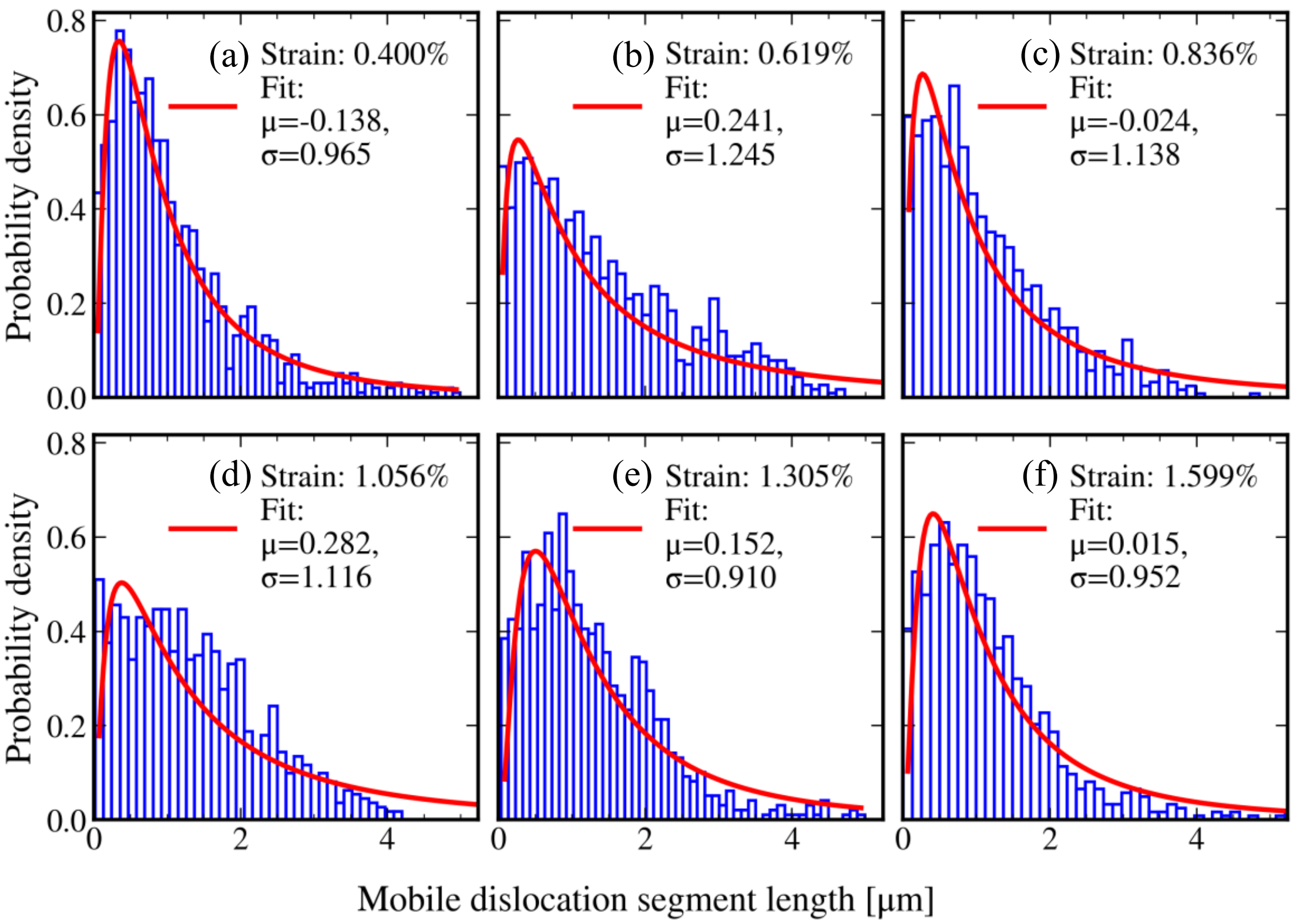

Figure 6. Distributions of mobile dislocation segment length at different strains for slip system $(111)[0\bar{1}1]$. The data is plotted together with the log-normal fits, and the corresponding distribution parameters are listed on top of each panel.

In order to investigate the stages of development of each slip system, and to observe the differences among them, Figure 7(a) plots the mean mobile segment length, $\mathrm{Mean}(l_{\mathrm{mobile}})$, as a function of strain for each of the 8 active slip systems. Each slip system exhibits an initial increase in the mean mobile segment length, followed by a decay regime, however, the peak location that separates both regimes differs per slip system, suggesting different rates of microstructural evolution. Figure 7(b) plots the average mobile dislocation segment length versus the inverse of the applied stress, which shows a positive correlation for the trend. These results demonstrate that the mean segment length decreases with increasing stress, following what appears to be a similitude law. Other similitude laws in plasticity can be found in the relatively recent review (Sauzay and Kubin, 2011), which are material invariant in FCC metals. Shorter mobile dislocation segment length implies that the length of dislocation source is also shorter. As a result, higher applied resolved shear stress is required to activate these dislocation sources (Balluffi, 2016; Hirth and Lothe, 1982).

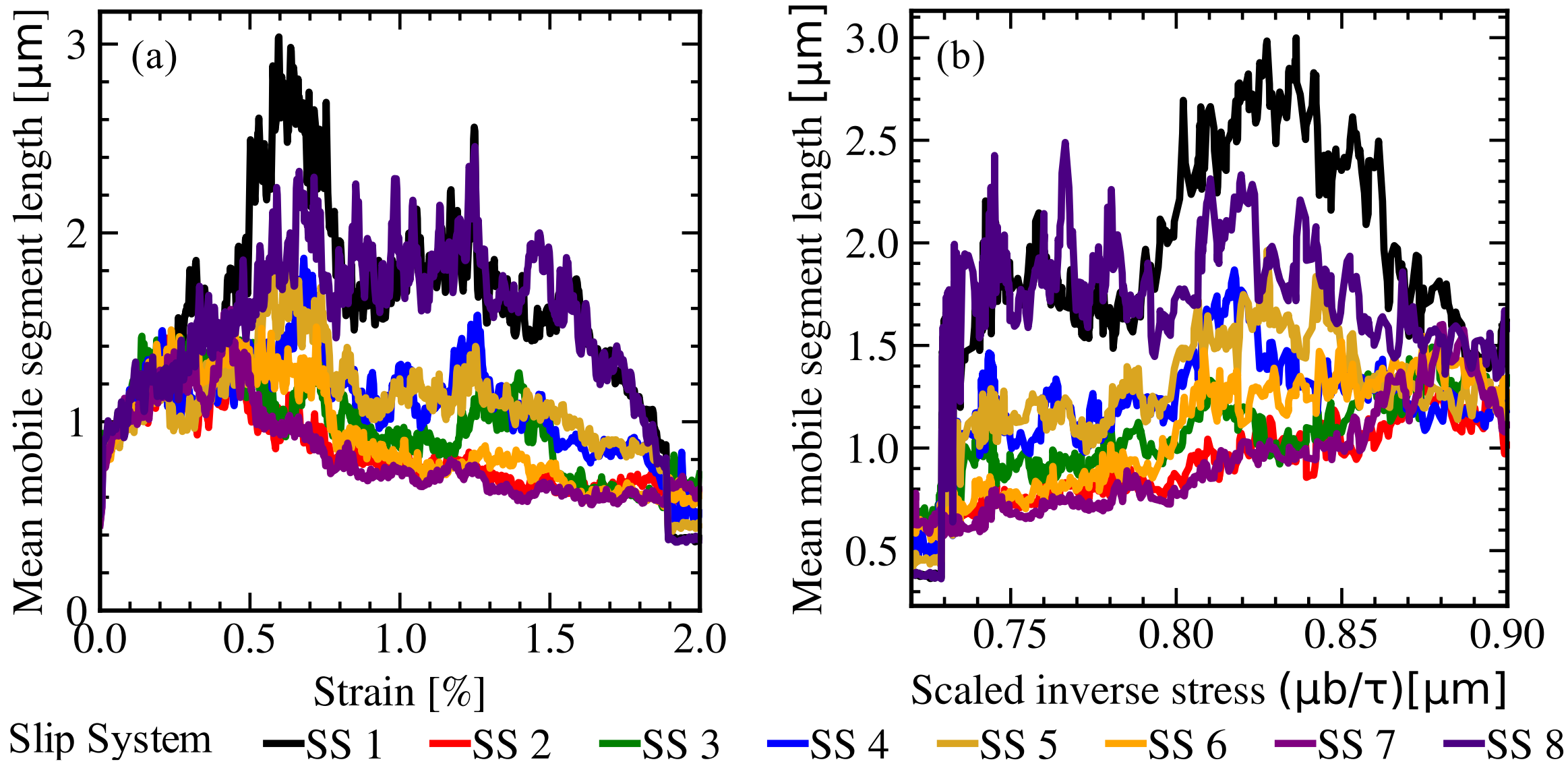

Figure 7. Mean mobile dislocation segment length versus (a) strain and (b) the scaled inverse of stress.

#### 4.2.2. Dislocation free path statistics

The mobile dislocations mean free path is another important length scale in microstructure-based crystal plasticity models since it controls the shear rates and the dislocation density evolution; basically, shorter mean free paths imply frequent obstructions that result in immobilization or division of mobile segments developing local dislocation pileups that require higher stress to unpin. In DDD, the mean free path can be determined by explicitly tracking the individual dislocations and their blockage by other dislocations via reactions at short range (Devincre et al., 2008). In CDD, the dislocations mean free path can be detected by inspecting the dislocation velocity field. Basically, the mean free path (or free glide distance) of a dislocation segment can be defined as the distance over which the velocity remains above a certain threshold. This distance can be analyzed using streamlines of the dislocation velocity field. In order to use the velocity field for this purpose, it must be monitored over the time the dislocation travels its free glide distance. However, it is possible to analyze the mean free path of the dislocations from a single snapshot of the velocity field if the velocity field does not change significantly over the time the mobile dislocations traverse their free glide distances. In order to ensure that this is the case, we define and compare two characteristic time scales, the "rise time" and the "travel time." The rise time is the time required for the dislocation status to change from immobile to mobile values at a specific location. The travel time is the time required for a mobile dislocation to travel along its mean free path. If the travel time is much smaller than the rise time, then the dislocation velocity field is said to evolve slowly, and the velocity field at a single snapshot of the dislocation configuration is adequate for computing the free path by searching for the distances between pinning points along the trajectory of the dislocation line (streamline of the velocity field).

Dislocation velocities at $N = 10^4$ random points in the simulation domain are monitored during the deformation to analyze the rise time and travel time. The dislocation velocity profiles of twenty sample points are shown in Figure 8. The dislocation velocities are sampled within strain intervals of $\pm 0.004\%$ at 0.404%, 0.412%, 1.003%, 1.019% 1.603% and 1.611% strain. For an applied strain rate of 20 $\mathrm{s}^{-1}$, the strain interval of 0.008% corresponds to $4 \mu\mathrm{s}$. The results show that velocity at some locations is above the threshold defining a mobile dislocation, with is $\mathrm{v}_{\mathrm{tol}} > 0.01$ m $\mathrm{s}^{-1}$. For example, five of the twenty points have dislocation velocity in the mobile range over the monitoring window 0.404% in Figure 8, while only two of the twenty points have a dislocation velocity in the mobile range over the monitoring window of 1.003%.

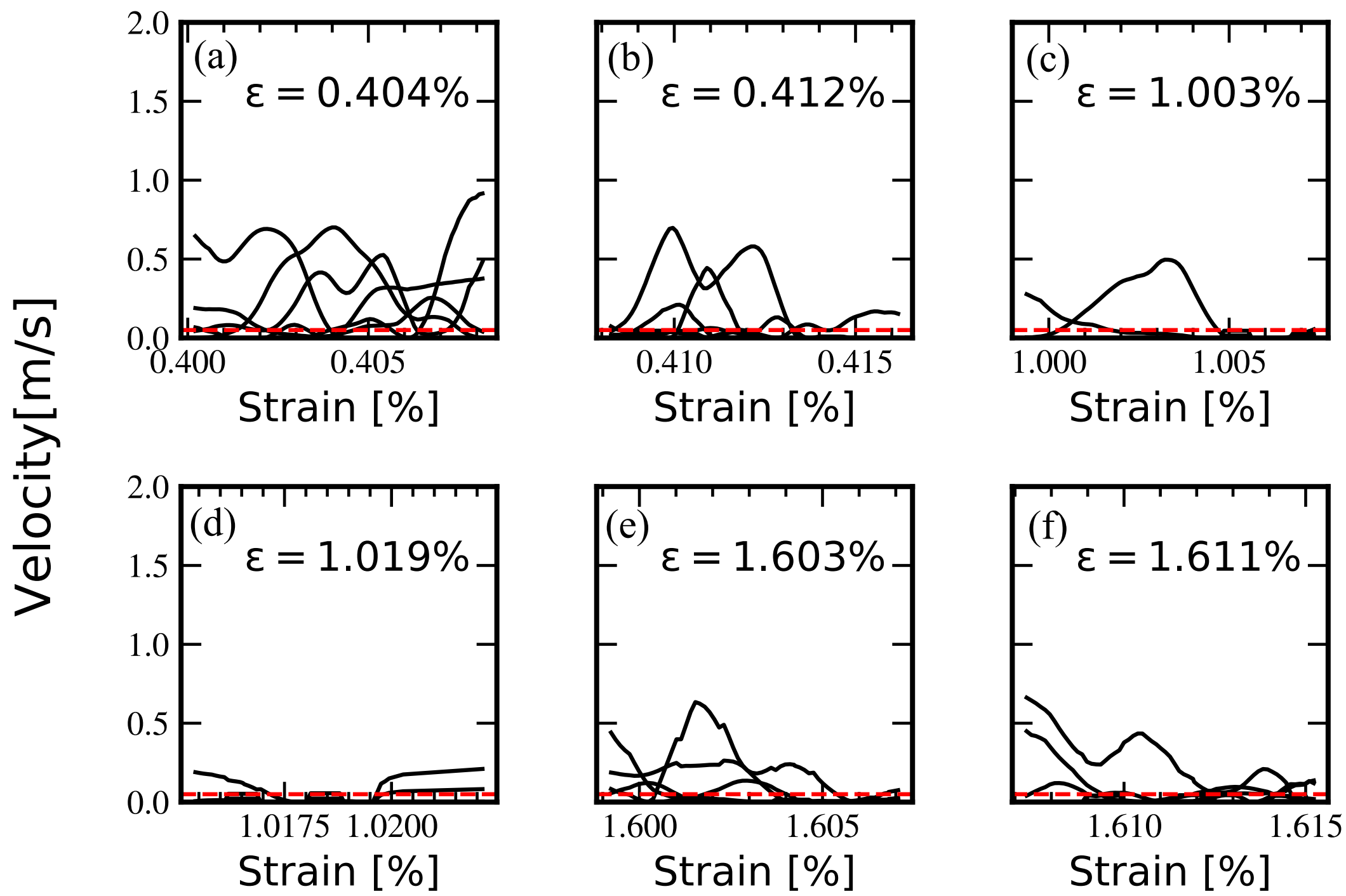

Figure 8. Evolution of the dislocation velocity at ten random points out of the $10^4$ random points analyzed in the simulations. The velocity data is collected within strain interval $\pm 0.004\%$ at (a) 0.404%; (b) 0.412%; (c) 1.003%; (d) 1.019%; (e) 1.019% (d) 1.603% (e) 1.611% strain levels. The velocity at some points remains zero over the sampling intervals at various strain levels.

To quantitatively analyze how fast the dislocation velocities are changing, the rise time at the $10^4$ sample points are calculated and its distribution at different strains are plotted in Figure 9. The vertical axis in this figure denotes fraction of points $n(t_{rise})/N$, where $n(t_{rise})$ is the number of points corresponding to a given rise time. We first pick the subset of all points with dislocation velocity less than the 0.01 m/s at each strain. Then we track the evolution of dislocation velocities at these points. A rise time is recorded when the dislocation velocity exceeds 0.01 m/s. If the dislocation velocity never exceeds 0.01 m/s over the monitoring interval, the rise time is recorded as $4 \times 10^3$ ns, which is the monitoring time itself. Figure 9 indicates that most of the initially immobile dislocations remain immobile within the monitoring window and even for the dislocations that eventually become mobile, the rise time is larger than $2 \times 10^3$ ns. As we will show later, the mean free path is in the order of 1.5 micrometers, while the average dislocation velocity along the mean free path is typically in the order of 6 m/s. Thus, the travel time will be in the order of 250 ns – a factor of 16 shorter than the rise time. Therefore, it is possible to use velocity field at a single snapshot of the dislocation configuration to compute the mean free path with the help of the velocity streamline analysis. We remark here that this finding is consistent with the nature of dislocation ensembles in deforming crystals, that a large fraction of the dislocation density is immobile, and that, on average, dislocations remain in an immobile state for much longer times than in a mobile state.

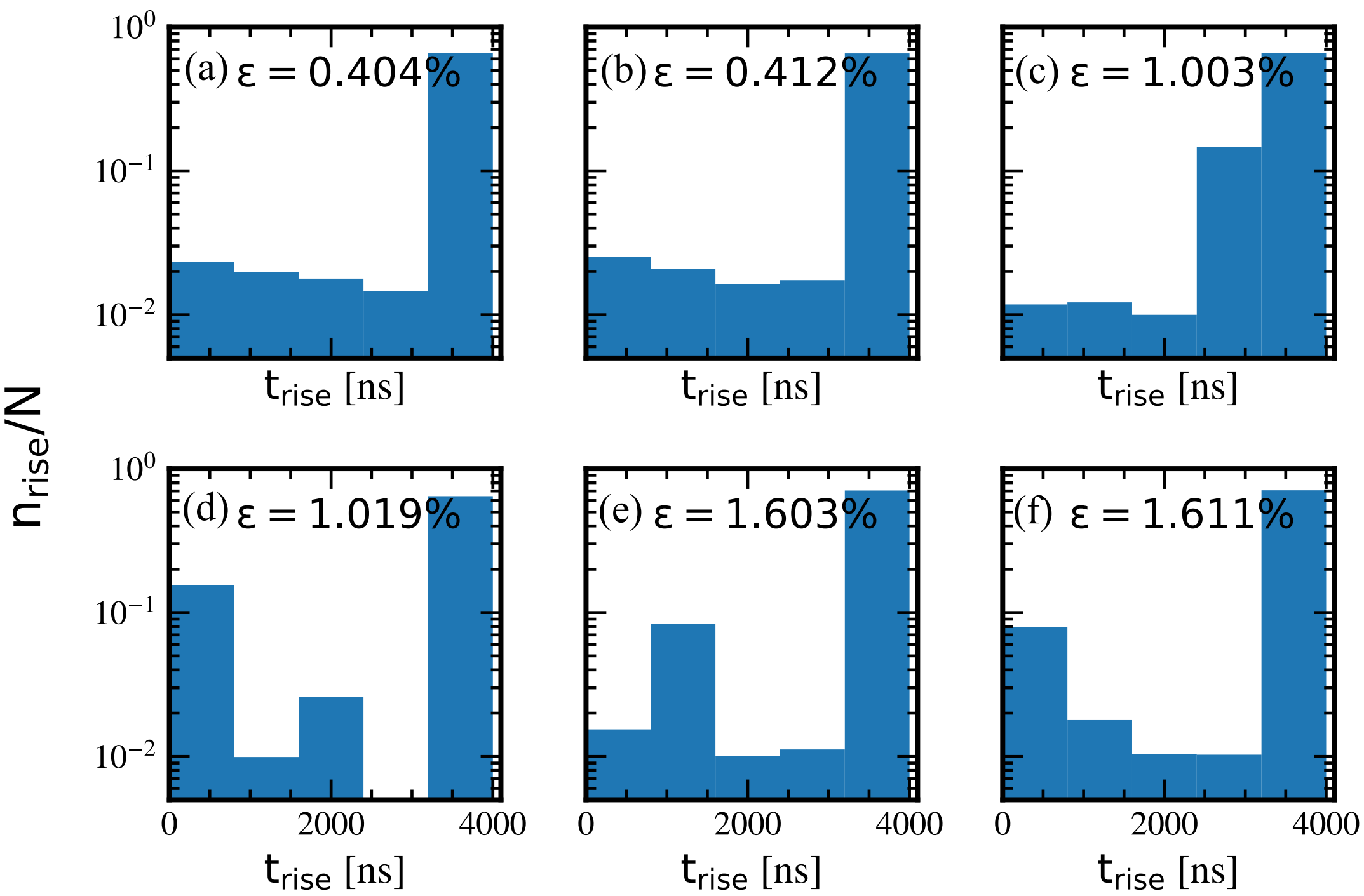

Figure 9. The histograms of rise time at $10^4$ random points in the simulation domain at (a) 0.404%; (b) 0.412%; (c) 1.003%; (d) 1.019%; (e) 1.603%; (d) 1.611% strain. The vertical axis records the fraction of points corresponding to each rise time.

We now proceed with the streamline analysis to compute the dislocation mean free path as discussed in Section 4.2.1. An adaptive Runge-Kutta method is used to solve Eq. (25) starting from $10^3$ initial seed points. The resulting streamlines of dislocation mean free path are shown in Figure 10. Here, only the streamlines for slip system $(111)[0\bar{1}1]$ at 0.4% and 1.6% strain are shown as an example. The color on each streamline represents the local dislocation velocity, which varies along the dislocation free path with highest values near the middle. Hence, the average dislocation velocity is approximated as half of the highest value, which was used to calculate the travel time.

Comparing Figure 10 and Figure 4, the streamlines of dislocation free path are both spatially correlated within the same spatial clusters. This is consistent with discrete dislocation dynamics (DDD) simulation results (Stricker et al., 2018; Stricker and Weygand, 2015), where mobile dislocations were shown to vary between 0.1 – 1 μm.

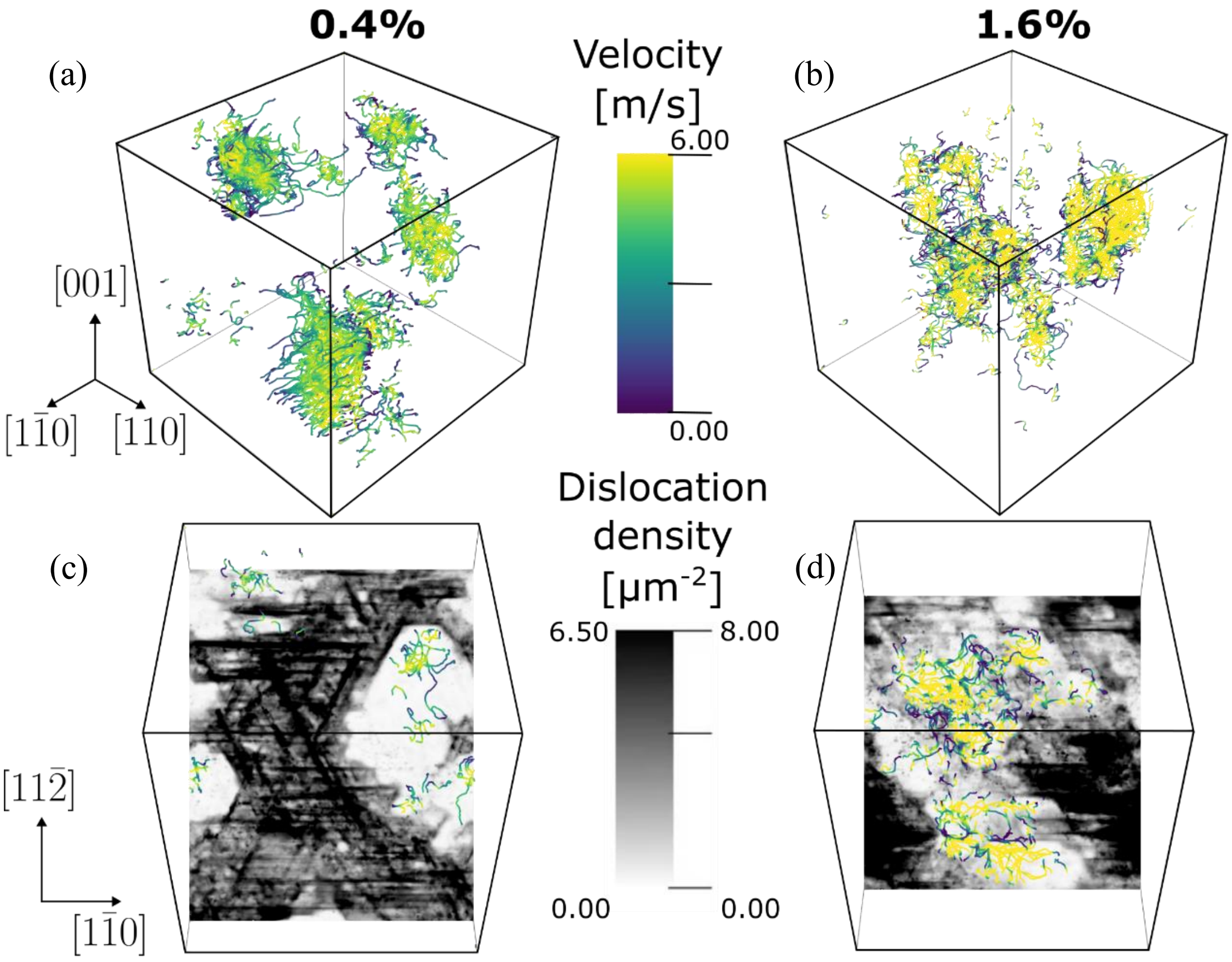

Figure 10. Streamlines of the free path for slip system $(111)[0\bar{1}1]$ at 0.4% (a,c) and 1.6% (b,d) strain. Panels (a,b) show velocity-colored segments for the full domain, while (c,d) overlay dislocation density on the $(111)$ plane, highlighting the free path correlation with cell interior regions.

The distributions of dislocation mean free path for the active slip systems at 1.0% strain are shown in Figure 11. It is estimated by first calculating the length of each segment along the dislocation velocity streamlines shown in Figure 10 and the local dislocation density at each point. The histograms are then plotted by counting the number of points on the free path, weighted by the dislocation density. As Figure 11 shows, there is variability of the distributions among active slip systems. Again, the distributions were fit to log-normal distributions, following Eq (26). The longest free path observed is slightly longer than 5 μm, which spreads over 70~75 mesh elements observed in slip system 1 and 8. The majority of mobile dislocations seem to have a free path falling in the range 1 μm spanning ~20 mesh elements.

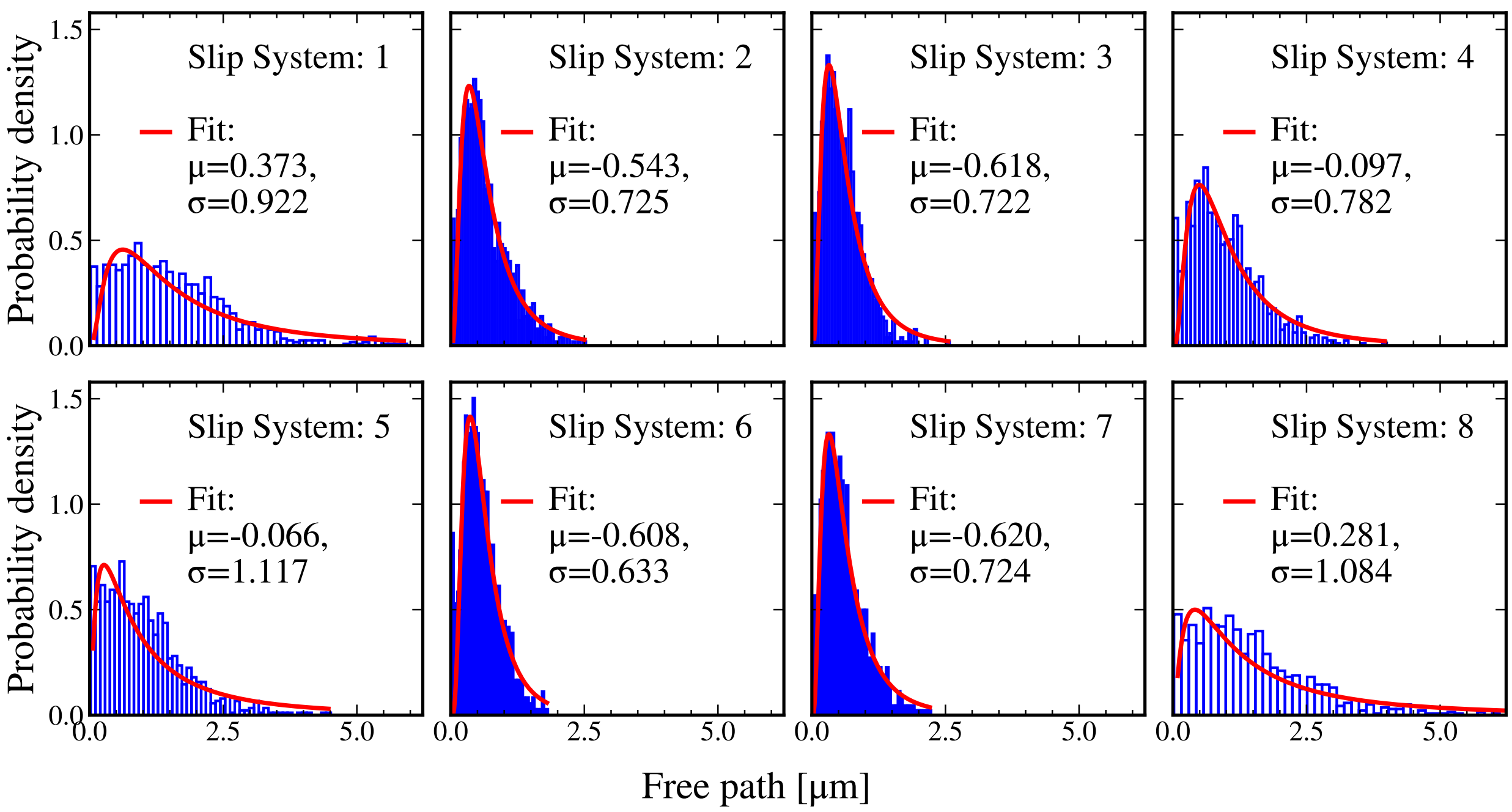

Figure 11. Distributions of dislocation free path distance for the eight active slip systems at 1.0% strain.

The distributions of dislocation free path at different strains (0.400%, 0.619%, 0.836%, 1.056%, 1.305%, and 1.599%) are shown in Figure 12 for slip system $(111)[0\bar{1}1]$. Across all strain levels, the distributions are strongly right-skewed and are well captured by lognormal fits shown with red lines with the fitting parameters added. The long tail of the distribution extends to 4-5 μm, while the modal free path lies well below 1μm at 0.400% strain, shifts to larger values and broadens around 1.056%, and then shifts back toward shorter path at higher strain 1.31-1.6%. This evolution trend is consistent with recent results captured using DDD simulations (Shimanek et al., 2025).

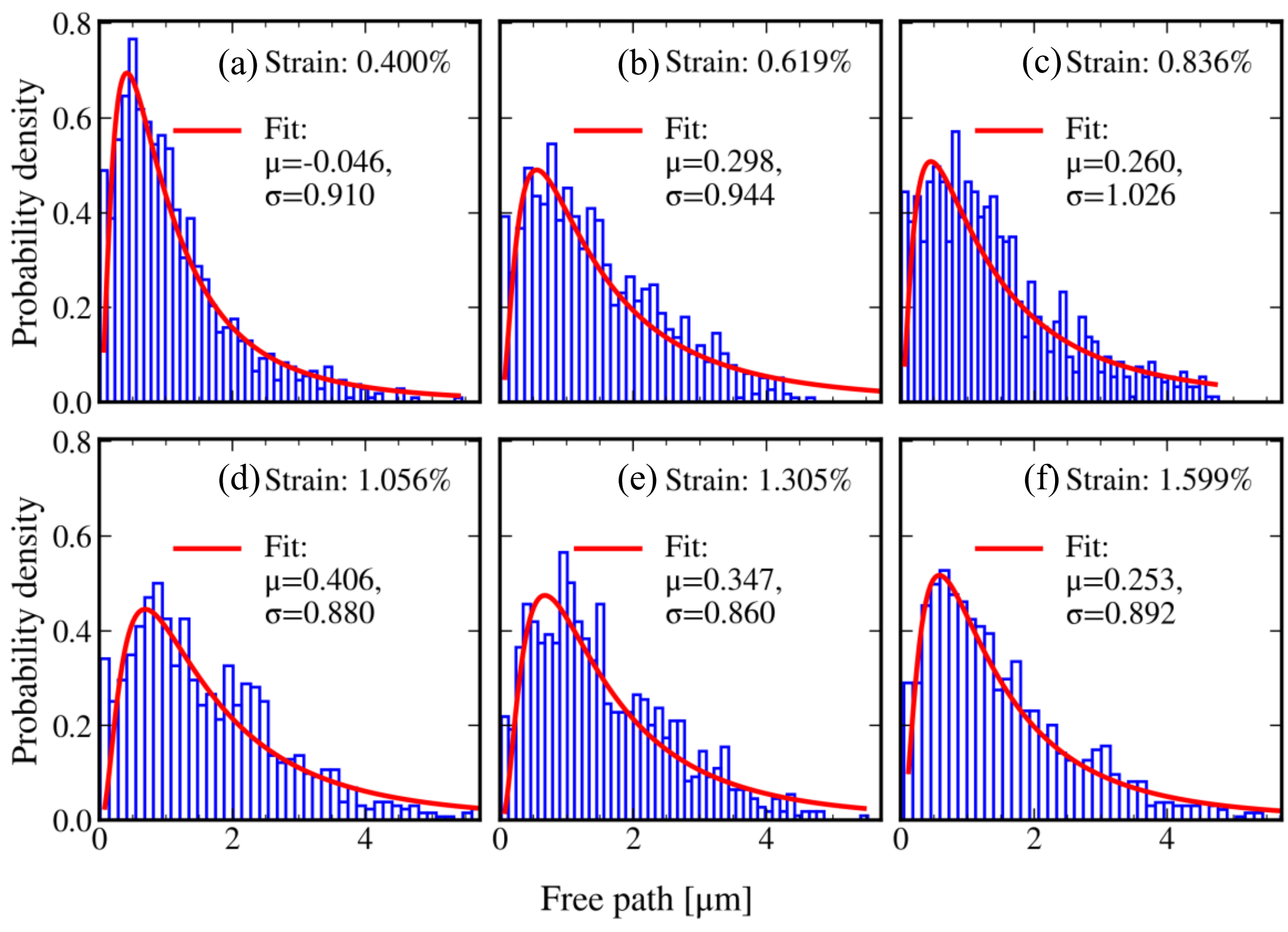

Figure 12. Distributions of dislocation free path length for slip system $(111)[0\bar{1}1]$ at different strains.

Figure 13(a) shows that the mean free path following the results in Figure 12; an initial increase up to 0.5-1.0% strain, then rapid decay up to 2.0% strain. The ratio of the mean free path to mobile dislocation segment length (η) is also plotted in Figure 13(b). The results in Figure 13(b) lead to an important conclusion regarding the formation of dislocation substructures which inform crystal plasticity models which dislocation microstructure is being formed and is used to calculate the mobile segment length. The evolution of this ratio maintains a mean value for all slip systems that is approximately equals to 1 but the values for the individual slip systems are increasingly deviating from 1. This means that the cells formed on individual slip systems may be slightly elongated but the geometric sum of cell structure on all slip systems approximates one that displays more equiaxed dimensions.

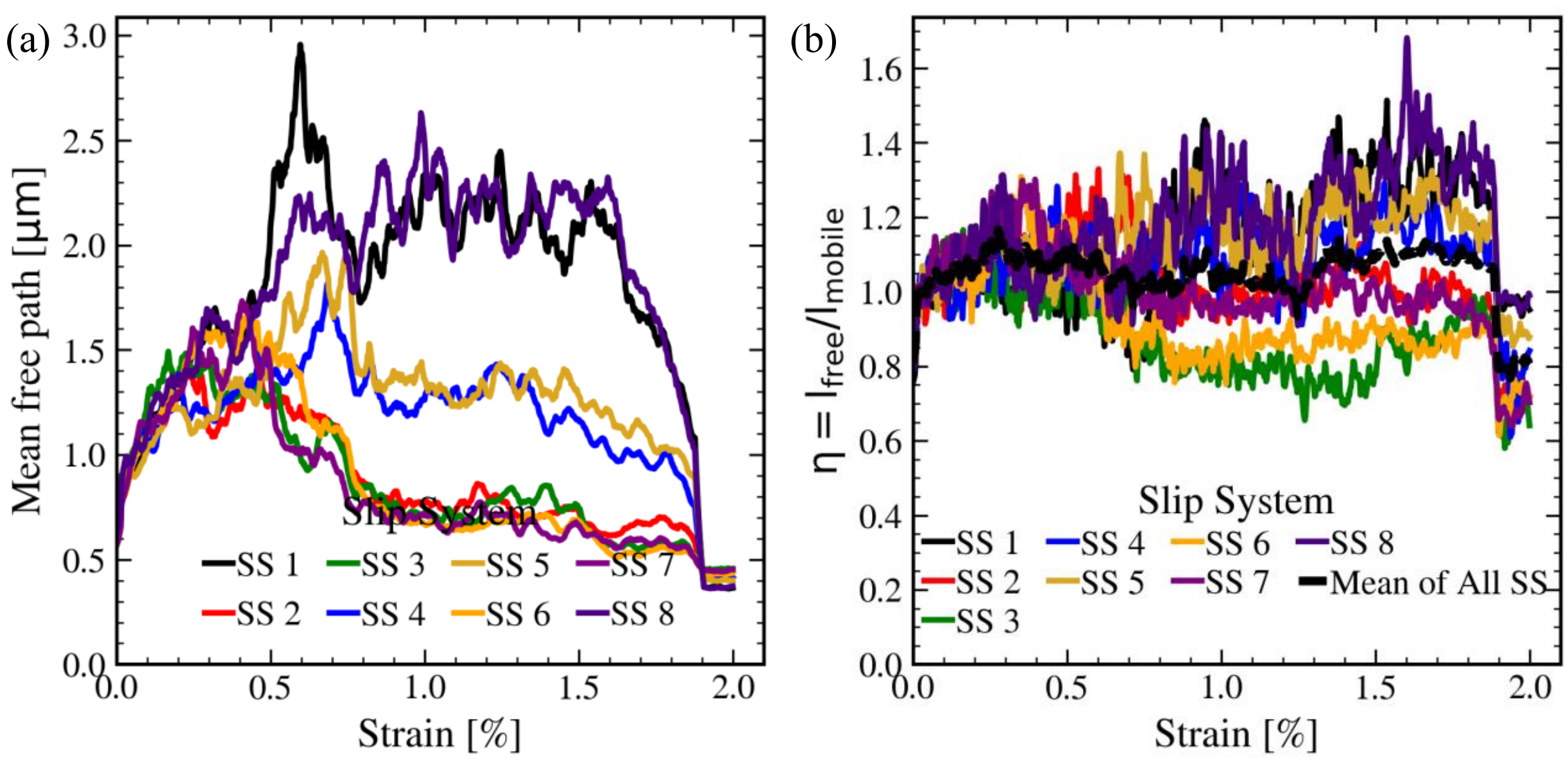

Figure 13. (a) Evolution of the dislocation mean free paths for active slip systems with strain. (b) The ratio $\eta = l_{free}/l_{mobile}$ versus strain.

#### 4.2.3. Dislocation wall volume fraction

Another characteristic distance associated with the dislocation substructure is the wall thickness, which, together with the structure dimensions, determines the wall volume fraction, $f_w$. Here, the wall volume fraction is calculated by assuming all immobile dislocations are constrained in dislocation walls. The dislocation wall volume is then obtained by summing the volume occupied by all immobile dislocations. The dislocation wall volume fraction versus strain is plotted in Figure 14(a) which forms high volume fraction of walls which is due to their broad structure that rapidly form in the early stages of deformation up to a peak value approaching 30% of the volume. At approximately 0.5%, the wall structures refine and stabilize accounting for much lower volume fraction represented by a rapid exponential decay beyond 0.5%. Pham and Holdsworth (2014) reported $f_w$ values in the range 0.1 to 0.7. As for the behavior of the wall volume fraction, the function form proposed by Estrin et al. (1998) for wall volume fraction evolution in Eq. (1) was found to fit the dislocation wall volume fraction CDD data when compared with the results in Figure 14(a), following an initial rise at very small strains.

The dislocation microstructure-based crystal plasticity models are based on the composite model proposed by Mughrabi (1983). In this model, the deformed crystal is assumed to consist of a relatively harder wall region of volume fraction $f_w$ and softer cells. A rule of mixture is then applied for both the over shear stress and shear strain rates in terms of their respective values in the two phases and the wall volume fraction (Castelluccio and McDowell, 2017; Dindarlou and Castelluccio, 2022). Inspection of the CDD results for the resolved shear stress distribution for slip system (1) at 0.4% strain within the cell and wall structures, shown in Figure 14(b), reveals the existence of two distinct stress populations, indeed. The wall structures exhibit a higher mean resolved shear stress with a moderately skewed distribution, whereas the cell interior display a markedly lower mean stress with a narrower distribution. The latter is captured quantitatively by the reduced scale parameter in the Cauchy distribution fit, despite the presence of secondary features at the distribution tails that have been previously discussed in the framework of the composite model by (Mughrabi, 1983). For further theoretical background, the reader is referred to (Gurtin, 2002; Mughrabi, 2001; Mughrabi and Ungár, 2002; Yefimov and Van Der Giessen, 2005).

The pronounced difference between the stress distributions in the wall and cell regions provides a measure of the back stress in crystal plasticity models, arising from the heterogeneous dislocation arrangement within the crystal. Within Mughrabi's composite model (Mughrabi, 1983), the channel resolved shear is expressed as,

$$\tau_c = \tau_a - f_w(\tau_w - \tau_c), \tag{27}$$

where $\tau_a$ is the applied resolved shear stress, $\tau_w$ is the wall stress, and $\tau_c$ is the channel stress (cell stress in the present case). The associated back stress is then defined as,

$$\tau_b = f_w(\tau_w - \tau_c), \tag{28}$$

which corresponds to the difference between the mean stresses of the wall and cell parts of the structure.

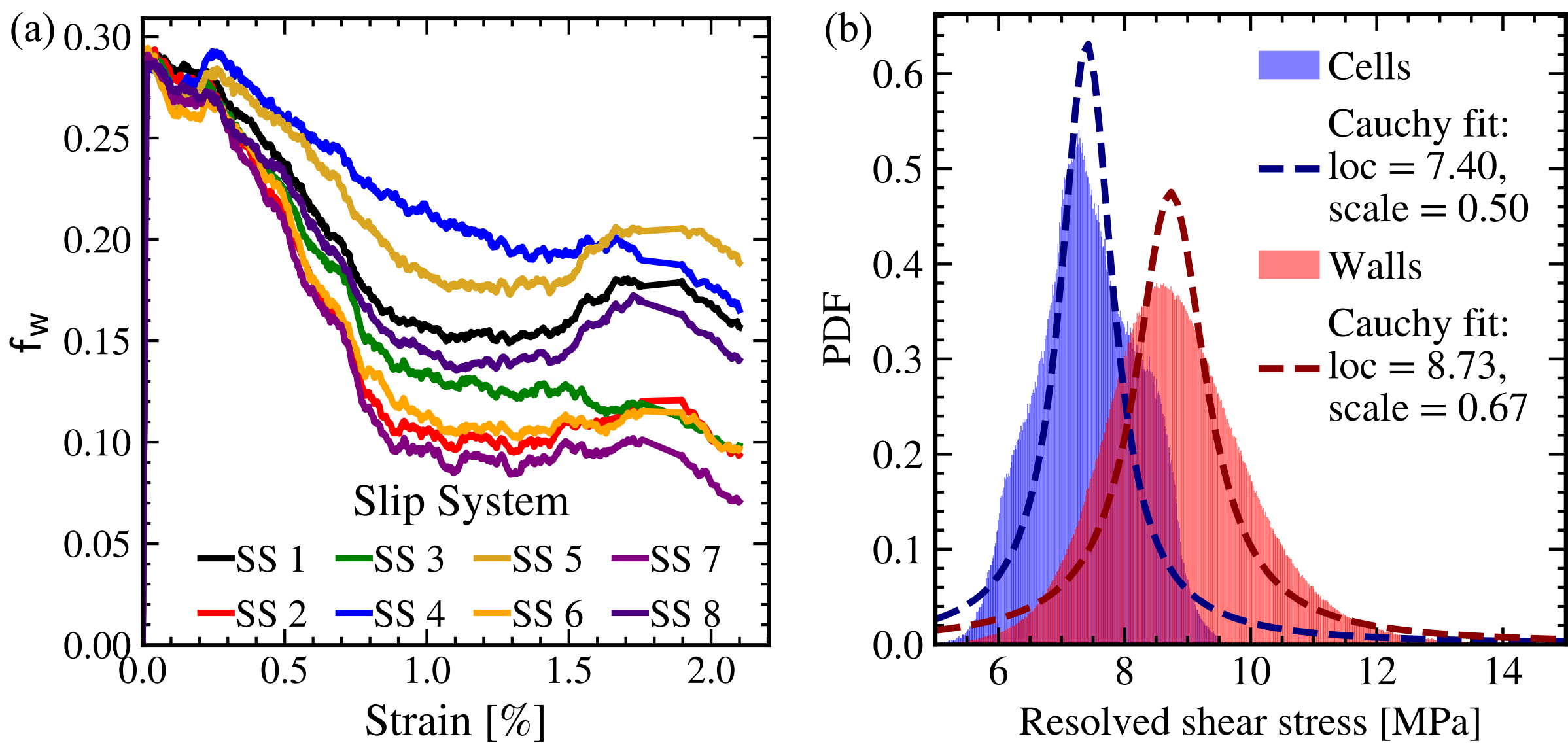

Figure 14. (a) Dislocation wall volume fraction versus strain, (b) resolved shear stress distribution in cell structures versus walls, both fitted to Cauchy distribution with fitting paramters listed in the legend for a single slip system (1) at 0.4% strain.

### 4.3. A short discussion

According to Dindarlou and Castelluccio (2022), the mean spacing of sessile dislocation walls, denoted $d_{struc}$, is approximated as the mean mobile dislocation length, $l_{mobile}$. This quantity follows the similitude relationship as reviewed in Sauzay and Kubin (2011), This relation is expressed in the famous similitude form:

$$d_{struc} = K_{struc} \frac{\mu b}{\max\limits_{k \in \mathcal{K}} |\tau^{(k)}|}, \tag{29}$$

where $\mu b / \max\limits_{k \in \mathcal{K}} |\tau^{(k)}|$ represents the scaled inverse stress and $K_{struc}$ is the similitude coefficient. The adoption of a single average value of $d_{struc}$ across all slip systems is a simplification that ignores the asymmetry of the dislocation self-organization on various slip systems, which is reflected in the slip system-specific values of the mobile segment length, $l_{mobile}$, shown in Figure 7. Additionally, in the microstructure-based crystal plasticity model of Dindarlou and Castelluccio (202), $\eta$ and $f_w$ are selected based on strain criteria from a menu of microstructures compiled from experimental observations. These two microstructure descriptors are again represented by one value each across all slip systems. Obviously, this is not the case for $\eta$ and $f_w$, as depicted in Figures 13b and 14a, respectively. As mentioned in the introduction, the mean free path is linked with the mean mobile length by $l_{free} = \eta l_{mobile}$, see Eq. (5). The assumption of a constant parameter $\eta$ linking the microstructural lengths $l_{free}$ and $l_{mobile}$ is supported by the trends in Figure 13b, namely that the average value over all slip systems is a constant close to unity. Considering that the slip system-specific values fall in the range $1 \pm 0.2$, this, again, supports the assumption that $l_{free} \approx l_{mobile}$ on average, for all slip systems. Given that $l_{mobile} \approx d_{struc}$, the three length scales can be considered equal to first order in the crystal plasticity model when cell structure is dominant.

With regard to the wall fraction, $f_w$, Figure 14a demonstrates pronounced slip system dependance, particularly at higher strains. This observation is consistent with the asymmetry in the density evolution of various slip systems, Figure 2d, and it warrants a bit of explanation here. In the literature on microstructure of plastically deformed metals, crystals are viewed in terms of cells and walls, or generally, regions of high and low dislocation densities. Reinforced mostly by TEM observations, this view does not account for the fact that individual slip systems may not be all participating in walls or high-density boundaries. We here adopt the modified view that the cellular structure may not cut across all slip systems. This modified view is actually supported by how TEM visualization works, that only dislocations satisfying a nontrivial $\mathbf{g} \cdot \mathbf{b}$ values ($\mathbf{g}$ and $\mathbf{b}$ being a reciprocal lattice vector and Burgers vector) can be viewed in a typical TEM micrograph (Williams and Carter, 2009). Our simulations clearly show that the tangled dislocations in a wall region do not generally include contributions from all slip systems. As such, the slip system values of $f_w$ are important to investigate. Averaging those values to have a single parameter for the entire crystal then becomes an assumption that the crystal plasticity modelers need to justify.

Next, it is important to explain the behavior of the wall volume fraction both across slip systems and for the individual ones. First, the variability of $f_w$ among slip systems may be tied to the slip rates and dislocation density evolution on various slip systems as shown in Figure 2c and Figure 2d. For a single slip system, however, we note that that $f_w$ exhibits a rapid initial rise followed by a short decay interval and a short rise interval, followed by a longer interval of decay that appears to be exponential before it slightly rises slowly one more time. This kind of behavior seems to be synchronous with how cells form in the system. As observed in the simulations, cells initially form with a large size in the domain then undergo restructuring, then reappear at smaller size in the regions defined by the intersections of the long boundaries (geometrically necessary boundaries, or GNBs) shown in Figure 3. When the local mean stress overcomes regions of high dislocation density, where tangling generates substantial back stress, it destabilizes the existing microstructure in the process. These regions act as nucleation sites for the formation of newer cells. The newly formed microstructure that appear in Figure 3c consists of smaller cells than the ones found in Figure 3b which are bounded by similar interfaces. Such boundaries can, in turn, serve as additional nucleation sites, enabling a recursive refinement of the microstructure. The very initial and rapid rise of $f_w$ seems to coincide with the short time over which the dislocations start to intersect each other. The short initial decay that follows this initial transient is associated with the appearance of large cells (size ~ half the simulation domain). These large cells are soon divided by more dislocation intersections giving rise to the values of $f_w$. The longer decay phase is associated with a more massive self-organization of dislocations into cells. This process takes more of the time (strain interval) before it slows down with the wall fractions gently rising again toward the end of the simulations due to a refinement process leading to creation of more walls as smaller cells form.

The current results show that all parameters identified in the microstructure-based crystal plasticity framework of the kind of Dindarlou and Castelluccio (2022) can be made slip system specific but, again, extensive data generation is required to support that. In this regard, CDD simulations of the kind presented here may be useful in generating those data. An important related issue that might want to be entertained by crystal plasticity model developer is the possibility of consideration of the full distributions of the microstructure parameters instead of their mean values. In our opinion, the crystal plasticity framework has the capacity to do so because all strain histories are resolved at the slip system level. However, that is a matter of tradeoff between the need to capture such details at the level of crystal plasticity versus speed and ability to actually generate the data from lower scale models and/or experiments.

## 5. Concluding remarks

Continuum dislocation dynamics is an effective method for simulating the evolution of dislocation substructures in crystalline materials. It has been applied to study dislocation pattern formation during deformation (Arora and Acharya, 2020; Groma et al., 2016; Sandfeld and Zaiser, 2015). However, most of these approaches are performed on two-dimensional dislocation ensembles. In the current work, the properties of 3D dislocation substructures are analyzed within the framework of dislocation structure-based crystal plasticity models, with a focus on the dislocation mobile segment length, free path and wall volume fraction. The vector density-based CDD method (Lin et al., 2021b; Lin and El-Azab, 2020; Xia and El-Azab, 2015) has been used to obtain the dislocation structure information. In this CDD method, dislocations on each slip system are represented by a vector density field, within the line bundle approximation of the dislocation configuration. The dislocation density vectors carry the information of both the scalar dislocation density and the line direction. The evolution of the dislocation density vectors is governed by a set of transport-reaction equations, coupled with the mechanical equilibrium equation. Finite element method is implemented to solve these equations.

Analyzing the three-dimensional dislocation substructure is challenging due to the curved nature of dislocation lines and their mutual interactions. A novel streamline method is proposed to obtain the characteristics of dislocation substructures generated by continuum dislocation dynamics. The streamlines can be established both for the dislocation density and the dislocation velocity vector fields. Streamlines are constructed by travelling along the given vector field starting at a suitably chosen set of initial points using an implicit Runge-Kutta method.

Substructure-sensitive crystal plasticity models rely on mesoscale parameterizations to quantify the underlying dislocation hardening mechanisms. Here, the streamline method is applied for a uniaxially loaded Ni crystal with eight active slip systems to analyze the evolution of the mobile dislocation segment length and dislocation free path for each slip system. The results obtained from our analysis showed that both the average mobile segment length and the dislocation mean free path initially increase and then decrease with increasing strain with high variability across the different slip systems. In addition, our analysis also revealed that the ratio of mean free path to mobile dislocation segment length evolves with strain, deviating from unity, both above and below, depending on the slip system. For a crystal oriented for loading along the [001] axis, the results have been successfully parameterized for the mobile dislocation segment length, the mobile dislocation mean free path, and the wall volume fraction. These results can be directly used to inform crystal plasticity models and support an effective strategy for bridging scale in bottom-up modelling approaches. Lastly, a connection between CDD and the composite model of Mughrabi (1983) was made, where the resolved shear stress in cells and walls was determined and found to be consistent with the basic assumption of that model. This builds confidence in the microstructure-sensitive crystal plasticity models and provides input to these models in regard to the back stress.

Overall, the current work demonstrates that the combination of CDD modeling and intelligent analysis of the dislocation microstructure data obtained with this method can indeed inform microstructure-sensitive macroscale crystal plasticity models. The streamline method developed in this work provides the data analysis tools required to make interface CDD and crystal plasticity. Although the applied strain is limited to relatively small values under monotonic loading, the characteristics of dislocation substructure have been captured and the capabilities of the approach have been demonstrated. Future work may extend this effort to larger strain and situations with variable loading paths, e.g., cyclic and ratchetting loading situations.

**Acknowledgements**

The authors are grateful for the support from the Naval Nuclear Laboratory, operated by Fluor Marine Propulsion, LLC for the US Naval Reactors Program. The initial development of the method was supported by the National Science Foundation, Division of Civil, Mechanical, and Manufacturing Innovation (CMMI), through award number 1663311 and by the US Department of Energy, Office of Science, Division of Materials Sciences and Engineering, through award number DE-SC0017718 at Purdue University. The results included in this publication and the final analysis and writing were all supported by the U. S. Department of Energy, Office of Fusion Energy Sciences, through award number DE-SC0024585 at Purdue University. During the final revisions of the manuscript, Peng Lin was supported by the Young Scientists Fund of the National Natural Science Foundation of China, Grant No. 12302129.

## References

Acharya, A., Roy, A., 2006. Size effects and idealized dislocation microstructure at small scales: Predictions of a Phenomenological model of Mesoscopic Field Dislocation Mechanics: Part I. J Mech Phys Solids 54, 1687–1710. https://doi.org/10.1016/J.JMPS.2006.01.009

Akhondzadeh, S., Sills, R.B., Bertin, N., Cai, W., 2020. Dislocation density-based plasticity model from massive discrete dislocation dynamics database. J Mech Phys Solids 145, 104152. https://doi.org/10.1016/J.JMPS.2020.104152

Akhondzadeh, Sh., Bertin, N., Sills, R.B., Cai, W., 2021. Slip-free multiplication and complexity of dislocation networks in FCC metals. Materials Theory 2021 5:1 5, 1–24. https://doi.org/10.1186/S41313-020-00024-Y

Anderson, J.P., Vivekanandan, V., Lin, P., Starkey, K., Pachaury, Y., El-Aza, A., 2022. Situating the Vector Density Approach among Contemporary Continuum Theories of Dislocation Dynamics. Journal of Engineering Materials and Technology, Transactions of the ASME 144. https://doi.org/10.1115/1.4052066/1115557

Arora, R., Acharya, A., 2020. Dislocation pattern formation in finite deformation crystal plasticity. Int J Solids Struct 184, 114–135. https://doi.org/10.1016/J.IJSOLSTR.2019.02.013

Arsenlis, A., Parks, D.M., Becker, R., Bulatov, V. V., 2004. On the evolution of crystallographic dislocation density in non-homogeneously deforming crystals. J Mech Phys Solids 52, 1213–1246. https://doi.org/10.1016/J.JMPS.2003.12.007

Asaro, R.J., Rice, J.R., 1977. Strain localization in ductile single crystals. J Mech Phys Solids 25, 309–338. https://doi.org/10.1016/0022-5096(77)90001-1

Ashraf, F., Castelluccio, G.M., 2021. A robust approach to parameterize dislocation glide energy barriers in FCC metals and alloys. J Mater Sci 56, 16491–16509. https://doi.org/10.1007/S10853-021-06376-1/METRICS

Baird, J.D., Gale, B., 1965. Attractive dislocation intersections and work hardening in metals. Philosophical Transactions of the Royal Society of London. Series A, Mathematical and Physical Sciences 257, 553–590. https://doi.org/10.1098/rsta.1965.0015

Balluffi, R.W., 2016. Introduction to Elasticity Theory for Crystal Defects: Second Edition. Introduction to Elasticity Theory for Crystal Defects: Second Edition 1–634. https://doi.org/10.1142/9939/SUPPL_FILE/9939_CHAP01.PDF

Barkia, B., Vallet, M., Tanguy, A., Auger, T., Héripré, E., 2025. New insights into microstructure evolution and deformation mechanisms in additively manufactured 316L stainless steel. Materials Science and Engineering: A 934, 148327. https://doi.org/10.1016/J.MSEA.2025.148327

Belytschko, T., Kam, W.K., Moran, B., Elkhodary, K.I., 2014. Beams and Shells. Nonlinear Finite Elements For Continua and Structures 535–596.

Bertin, N., Bulatov, V. V., Zhou, F., 2024. Learning dislocation dynamics mobility laws from large-scale MD simulations. NPJ Comput Mater 10, 1–9. https://doi.org/10.1038/S41524-024-01378-4;SUBJMETA=1023,1026,1034,1037,301,303,639;KWRD=COMPUTATIONAL+METHODS,MECHANICAL+PROPERTIES,METALS+AND+ALLOYS

Bishara, D., Xie, Y., Liu, W.K., Li, S., 2022. A State-of-the-Art Review on Machine Learning-Based Multiscale Modeling, Simulation, Homogenization and Design of Materials. Archives of Computational Methods in Engineering 2022 30:1 30, 191–222. https://doi.org/10.1007/S11831-022-09795-8

Castelluccio, G.M., Geller, C.B., McDowell, D.L., 2018. A rationale for modeling hydrogen effects on plastic deformation across scales in FCC metals. Int J Plast 111, 72–84. https://doi.org/10.1016/J.IJPLAS.2018.07.009

Castelluccio, G.M., McDowell, D.L., 2017. Mesoscale cyclic crystal plasticity with dislocation substructures. Int J Plast 98, 1–26. https://doi.org/10.1016/J.IJPLAS.2017.06.002

Csikor, F.F., Motz, C., Weygand, D., Zaiser, M., Zapperi, S., 2007. Dislocation avalanches, strain bursts, and the problem of plastic forming at the micrometer scale. Science (1979) 318, 251–254. https://doi.org/10.1126/SCIENCE.1143719/SUPPL_FILE/CSIKOR.SOM.PDF

Cui, Y.N., Lin, P., Liu, Z.L., Zhuang, Z., 2014. Theoretical and numerical investigations of single arm dislocation source controlled plastic flow in FCC micropillars. Int J Plast 55, 279–292. https://doi.org/10.1016/J.IJPLAS.2013.11.011

Deng, J., El-Azab, A., 2010. Temporal statistics and coarse graining of dislocation ensembles. Philosophical Magazine 90, 3651–3678. https://doi.org/10.1080/14786435.2010.497472;WGROUP:STRING:PUBLICATION

Devincre, B., Hoc, T., Kubin, L., 2008. Dislocation mean free paths and strain hardening of crystals. Science (1979) 320, 1745–1748. https://doi.org/10.1126/SCIENCE.1156101/SUPPL_FILE/DEVINCRE.SOM.PDF

Devincre, B., Kubin, L., Hoc, T., 2006. Physical analyses of crystal plasticity by DD simulations. Scr Mater 54, 741–746. https://doi.org/10.1016/J.SCRIPTAMAT.2005.10.066

Dindarlou, S., Castelluccio, G.M., 2022. Substructure-sensitive crystal plasticity with material-invariant parameters. Int J Plast 155, 103306. https://doi.org/10.1016/j.ijplas.2022.103306

El-Azab, A., 2000. Statistical mechanics treatment of the evolution of dislocation distributions in single crystals. Phys Rev B 61, 11956. https://doi.org/10.1103/PhysRevB.61.11956

El-Azab, A., Po, G., 2018. Continuum Dislocation Dynamics: Classical Theory and Contemporary Models. Handbook of Materials Modeling 1–25. https://doi.org/10.1007/978-3-319-42913-7_18-1

Estrin, Y., Tóth, L.S., Molinari, A., Bréchet, Y., 1998. A dislocation-based model for all hardening stages in large strain deformation. Acta Mater 46, 5509–5522. https://doi.org/10.1016/S1359-6454(98)00196-7

Franciosi, P., Berveiller, M., Zaoui, A., 1980. Latent hardening in copper and aluminium single crystals. Acta Metallurgica 28, 273–283. https://doi.org/10.1016/0001-6160(80)90162-5

Franciosi, P., Zaoui, A., 1982. Multislip in f.c.c. crystals a theoretical approach compared with experimental data. Acta Metallurgica 30, 1627–1637. https://doi.org/10.1016/0001-6160(82)90184-5

Grilli, N., Janssens, K.G.F., Nellessen, J., Sandlöbes, S., Raabe, D., 2018. Multiple slip dislocation patterning in a dislocation-based crystal plasticity finite element method. Int J Plast 100, 104–121. https://doi.org/10.1016/J.IJPLAS.2017.09.015

Groh, S., Zbib, H.M., 2009. Advances in Discrete Dislocations Dynamics and Multiscale Modeling. https://doi.org/10.1115/1.3183783

Groma, I., Zaiser, M., Ispánovity, P.D., 2016. Dislocation patterning in a two-dimensional continuum theory of dislocations. Phys Rev B 93, 214110. https://doi.org/10.1103/PHYSREVB.93.214110/FIGURES/3/MEDIUM

Gruber, P.A., Böhm, J., Onuseit, F., Wanner, A., Spolenak, R., Arzt, E., 2008. Size effects on yield strength and strain hardening for ultra-thin Cu films with and without passivation: A study by synchrotron and bulge test techniques. Acta Mater 56, 2318–2335. https://doi.org/10.1016/J.ACTAMAT.2008.01.027

Gu, Y., Stiles, C.D., El-Awady, J.A., 2024. A statistical perspective for predicting the strength of metals: Revisiting the Hall–Petch relationship using machine learning. Acta Mater 266, 119631. https://doi.org/10.1016/J.ACTAMAT.2023.119631

Gurtin, M.E., 2002. A gradient theory of single-crystal viscoplasticity that accounts for geometrically necessary dislocations. J Mech Phys Solids 50, 5–32. https://doi.org/10.1016/S0022-5096(01)00104-1

Hansen, N., Huang, X., 1998. Microstructure and flow stress of polycrystals and single crystals. Acta Mater 46, 1827–1836. https://doi.org/10.1016/S1359-6454(97)00365-0

Hansen, N., Huang, X., Pantleon, W., Winther, G., 2006. Grain orientation and dislocation patterns. Philosophical Magazine 86, 3981–3994. https://doi.org/10.1080/14786430600654446;WGROUP:STRING:PUBLICATION

Hirth, J.Price., Lothe, Jens., 1982. Theory of dislocations 857.

Hochrainer, T., 2016. Thermodynamically consistent continuum dislocation dynamics. J Mech Phys Solids 88, 12–22. https://doi.org/10.1016/J.JMPS.2015.12.015

Hochrainer, T., 2015. Multipole expansion of continuum dislocations dynamics in terms of alignment tensors. Philosophical Magazine 95, 1321–1367. https://doi.org/10.1080/14786435.2015.1026297;CTYPE:STRING:JOURNAL

Hull, D., Bacon, D.J., 2011. Introduction to Dislocations, Fifth edition. ed. Elsevier. https://doi.org/10.1016/C2009-0-64358-0

Jiang, B., 1998. The Least-Squares Finite Element Method. Scientific Computation. https://doi.org/10.1007/978-3-662-03740-9

Jiang, J., Benjamin Britton, T., Wilkinson, A.J., 2015. Evolution of intragranular stresses and dislocation densities during cyclic deformation of polycrystalline copper. Acta Mater 94, 193–204. https://doi.org/10.1016/J.ACTAMAT.2015.04.031

Keller, R.M., Baker, S.P., Arzt, E., 1999. Stress–temperature behavior of unpassivated thin copper films. Acta Mater 47, 415–426. https://doi.org/10.1016/S1359-6454(98)00387-5

Kocks, U.F., Mecking, H., 2003. Physics and phenomenology of strain hardening: the FCC case. Prog Mater Sci 48, 171–273. https://doi.org/10.1016/S0079-6425(02)00003-8

Kosevich, A.M., 1965. DYNAMICAL THEORY OF DISLOCATIONS. Soviet Physics Uspekhi 7, 837. https://doi.org/10.1070/PU1965V007N06ABEH003688

Kröner, E., 1959. Allgemeine Kontinuumstheorie der Versetzungen und Eigenspannungen. Arch Ration Mech Anal 4, 273–334. https://doi.org/10.1007/BF00281393/METRICS

Kubin, L., Devincre, B., Hoc, T., 2008. Toward a physical model for strain hardening in fcc crystals. Materials Science and Engineering: A 483–484, 19–24. https://doi.org/10.1016/J.MSEA.2007.01.167

Lavenstein, S., El-Awady, J.A., 2019. Micro-scale fatigue mechanisms in metals: Insights gained from small-scale experiments and discrete dislocation dynamics simulations. Curr Opin Solid State Mater Sci 23, 100765. https://doi.org/10.1016/J.COSSMS.2019.07.004

Lax, P.D., 1973. Hyperbolic Difference Equations: A Review of the Courant-Friedrichs-Lewy Paper in the Light of Recent Developments. IBM J Res Dev 11, 235–238. https://doi.org/10.1147/RD.112.0235

Leung, H.S., Leung, P.S.S., Cheng, B., Ngan, A.H.W., 2015. A new dislocation-density-function dynamics scheme for computational crystal plasticity by explicit consideration of dislocation elastic interactions. Int J Plast 67, 1–25. https://doi.org/10.1016/J.IJPLAS.2014.09.009

Lin, P., El-Azab, A., 2020. Implementation of annihilation and junction reactions in vector density-based continuum dislocation dynamics. Model Simul Mat Sci Eng 28, 045003. https://doi.org/10.1088/1361-651X/AB7D90

Lin, P., Vivekanandan, V., Anglin, B., Geller, C., El-Azab, A., 2021a. Incorporating point defect generation due to jog formation into the vector density-based continuum dislocation dynamics approach. J Mech Phys Solids 156, 104609. https://doi.org/10.1016/J.JMPS.2021.104609

Lin, P., Vivekanandan, V., Starkey, K., Anglin, B., Geller, C., El-Azab, A., 2021b. On the computational solution of vector-density based continuum dislocation dynamics models:

A comparison of two plastic distortion and stress update algorithms. Int J Plast 138, 102943. https://doi.org/10.1016/J.IJPLAS.2021.102943

Monavari, M., Zaiser, M., 2018. Annihilation and sources in continuum dislocation dynamics. Materials Theory 2, 1–30. https://doi.org/10.1186/S41313-018-0010-Z/FIGURES/15

Mughrabi, H., 2001. On the role of strain gradients and long-range internal stresses in the composite model of crystal plasticity. Materials Science and Engineering: A 317, 171–180. https://doi.org/10.1016/S0921-5093(01)01173-X

Mughrabi, H., 1983. Dislocation wall and cell structures and long-range internal stresses in deformed metal crystals. Acta Metallurgica 31, 1367–1379. https://doi.org/10.1016/0001-6160(83)90007-X

Mughrabi, H., Ungár, T., 2002. Chapter 60 Long-Range internal stresses in deformed single-phase materials: The composite model and its consequences. Dislocations in Solids 11, 343–411. https://doi.org/10.1016/S1572-4859(02)80011-0

Mura, 1963. Continuous distribution of moving dislocations. Philosophical Magazine 8, 843–857. https://doi.org/10.1080/14786436308213841

Nye, J.F., 1953. Some geometrical relations in dislocated crystals. Acta Metallurgica 1, 153–162. https://doi.org/10.1016/0001-6160(53)90054-6

Olmsted, D.L., Hector, L.G., Curtin, W.A., Clifton, R.J., 2005. Atomistic simulations of dislocation mobility in Al, Ni and Al/Mg alloys. Model Simul Mat Sci Eng 13, 371. https://doi.org/10.1088/0965-0393/13/3/007

Peach, M., Koehler, J.S., 1950. The Forces Exerted on Dislocations and the Stress Fields Produced by Them. Physical Review 80, 436. https://doi.org/10.1103/PhysRev.80.436

Pham, M.S., Holdsworth, S.R., 2014. Evolution of relationships between dislocation microstructures and internal stresses of AISI 316L during cyclic loading at 293 K and 573 K (20 °c and 300 °c). Metall Mater Trans A Phys Metall Mater Sci 45, 738–751. https://doi.org/10.1007/S11661-013-1981-7/FIGURES/14

Press, W.H., Teukolsky, S.A., Vetterling, W.T., Flannery, B.P., 2007. Numerical Recipes 3rd Edition NUMERICAL RECIPES Third Edition.

Rafiei, M.H., Gu, Y., El-Awady, J.A., 2020. Machine Learning of Dislocation-Induced Stress Fields and Interaction Forces. JOM 72, 4380–4392. https://doi.org/10.1007/S11837-020-04389-W/FIGURES/7

Reuber, C., Eisenlohr, P., Roters, F., Raabe, D., 2014. Dislocation density distribution around an indent in single-crystalline nickel: Comparing nonlocal crystal plasticity finite-element predictions with experiments. Acta Mater 71, 333–348. https://doi.org/10.1016/J.ACTAMAT.2014.03.012

Roy, A., Acharya, A., 2006. Size effects and idealized dislocation microstructure at small scales: Predictions of a Phenomenological model of Mesoscopic Field Dislocation Mechanics: Part II. J Mech Phys Solids 54, 1711–1743. https://doi.org/10.1016/J.JMPS.2006.01.012

Sandfeld, S., Zaiser, M., 2015. Pattern formation in a minimal model of continuum dislocation plasticity. Model Simul Mat Sci Eng 23, 065005. https://doi.org/10.1088/0965-0393/23/6/065005

Sauzay, M., 2008. Analytical modelling of intragranular backstresses due to deformation induced dislocation microstructures. Int J Plast 24, 727–745. https://doi.org/10.1016/J.IJPLAS.2007.07.004

Sauzay, M., Kubin, L.P., 2011. Scaling laws for dislocation microstructures in monotonic and cyclic deformation of fcc metals. Prog Mater Sci 56, 725–784. https://doi.org/10.1016/J.PMATSCI.2011.01.006

Schoeck, G., Frydman, R., 1972. The Contribution of the Dislocation Forest to the Flow Stress. physica status solidi (b) 53, 661–673. https://doi.org/10.1002/PSSB.2220530227

SharafEldin, K., Miller, B.D., Liu, W., Tischler, J., Anglin, B., El-Azab, A., 2025. Informed Unsupervised Machine Learning Analysis of Dislocation Microstructure from High-

Resolution Differential Aperture X-Ray Structural Microscopy Data. https://doi.org/10.2139/SSRN.5257211

Shimanek, J.D., Bamney, D., Capolungo, L., Liu, Z.-K., Beese, A.M., 2025. Effect of anisotropic Peierls barrier on the evolution of discrete dislocation networks in Ni. Model Simul Mat Sci Eng 33, 025015. https://doi.org/10.1088/1361-651X/adad8f

Shishvan, S.S., Van der Giessen, E., 2010. Distribution of dislocation source length and the size dependent yield strength in freestanding thin films. J Mech Phys Solids 58, 678–695. https://doi.org/10.1016/J.JMPS.2010.02.011

Song, H., Gunkelmann, N., Po, G., Sandfeld, S., 2021. Data-mining of dislocation microstructures: concepts for coarse-graining of internal energies. Model Simul Mat Sci Eng 29, 035005. https://doi.org/10.1088/1361-651X/ABDC6B

Steinberger, D., Song, H., Sandfeld, S., 2019. Machine learning-based classification of dislocation microstructures. Front Mater 6, 452769. https://doi.org/10.3389/FMATS.2019.00141/BIBTEX

Stricker, M., Sudmanns, M., Schulz, K., Hochrainer, T., Weygand, D., 2018. Dislocation multiplication in stage II deformation of fcc multi-slip single crystals. J Mech Phys Solids 119, 319–333. https://doi.org/10.1016/J.JMPS.2018.07.003

Stricker, M., Weygand, D., 2015. Dislocation multiplication mechanisms – Glissile junctions and their role on the plastic deformation at the microscale. Acta Mater 99, 130–139. https://doi.org/10.1016/J.ACTAMAT.2015.07.073

Vivekanandan, V., Anglin, B., El-Azab, A., 2023. A data driven approach for cross-slip modelling in continuum dislocation dynamics. Int J Plast 164, 103597. https://doi.org/10.1016/J.IJPLAS.2023.103597

Vivekanandan, V., Lin, P., Winther, G., El-Azab, A., 2021. On the implementation of dislocation reactions in continuum dislocation dynamics modeling of mesoscale plasticity. J Mech Phys Solids 149, 104327. https://doi.org/10.1016/J.JMPS.2021.104327

Vivekanandan, V., Pierre Anderson, J., Pachaury, Y., Mohamed, M.S., El-Azab, A., 2022. Statistics of internal stress fluctuations in dislocated crystals and relevance to density-based dislocation dynamics models. Model Simul Mat Sci Eng 30, 045007. https://doi.org/10.1088/1361-651X/AC5DCF

Williams, D.B., Carter, C.B., 2009. Transmission electron microscopy: A textbook for materials science. Transmission Electron Microscopy: A Textbook for Materials Science 1–760. https://doi.org/10.1007/978-0-387-76501-3/COVER

Wu, R., Zaiser, M., 2021. Cell structure formation in a two-dimensional density-based dislocation dynamics model. Materials Theory 5, 1–22. https://doi.org/10.1186/S41313-020-00025-X/FIGURES/8

Xia, S., Belak, J., El-Azab, A., 2016. The discrete-continuum connection in dislocation dynamics: I. Time coarse graining of cross slip. Model Simul Mat Sci Eng 24, 075007. https://doi.org/10.1088/0965-0393/24/7/075007

Xia, S., El-Azab, A., 2015. Computational modelling of mesoscale dislocation patterning and plastic deformation of single crystals. Model Simul Mat Sci Eng 23, 055009. https://doi.org/10.1088/0965-0393/23/5/055009

Yang, Z., Papanikolaou, S., Reid, A.C.E., Liao, W. keng, Choudhary, A.N., Campbell, C., Agrawal, A., 2020. Learning to Predict Crystal Plasticity at the Nanoscale: Deep Residual Networks and Size Effects in Uniaxial Compression Discrete Dislocation Simulations. Sci Rep 10, 1–14. https://doi.org/10.1038/S41598-020-65157-Z;SUBJMETA=1034,1037,12,301,639,930;KWRD=CHARACTERIZATION+AND+ANALYTICAL+TECHNIQUES,COMPUTATIONAL+METHODS

Yefimov, S., Van Der Giessen, E., 2005. Multiple slip in a strain-gradient plasticity model motivated by a statistical-mechanics description of dislocations. Int J Solids Struct 42, 3375–3394. https://doi.org/10.1016/J.IJSOLSTR.2004.10.025